\documentclass[aoas]{imsart}

\RequirePackage{amsthm,amsmath,amsfonts,amssymb}
\RequirePackage[authoryear]{natbib}
\RequirePackage[colorlinks,citecolor=blue,urlcolor=blue]{hyperref}
\RequirePackage{graphicx}
\usepackage{multirow}
\usepackage{soul}

\startlocaldefs
\theoremstyle{plain}

\theoremstyle{remark}

\usepackage{todonotes}
\newcommand{\F}{\mathcal{F}}
\newcommand{\N}{\mathcal{N}}
\newcommand{\V}{\mathcal{V}}
\newcommand{\CRP}{\mathrm{CRP}}
\newcommand{\MCRP}{\mathrm{MCRP}}
\newcommand{\MDP}{\mathrm{MDP}}

\newcommand{\C}{\mathcal{C}}
\newcommand{\A}{\mathcal{A}}
\newcommand{\T}{\mathcal{T}}
\newcommand{\U}{\mathcal{U}}

\def\B{\text{B}}
\newcommand{\M}{\mathcal{M}}

 \def\X{\mathbf{X}}
\def\Y{\mathbf{Y}}
\mathchardef\mhyphen="2D 

\endlocaldefs

\begin{document}

\begin{frontmatter}
\title{Biclustering random matrix partitions with an application to classification of forensic body fluids}
\runtitle{Biclustering for classification}

\begin{aug}
\author[A]{\fnms{Chieh-Hsi}~\snm{Wu}\ead[label=e1]{c-h.wu@soton.ac.uk}\orcid{0000-0001-9386-725X}},
\author[B]{\fnms{Amy D.}~\snm{Roeder}\ead[label=e2]{aroeder@cellmark.co.uk}}
\and
\author[C]{\fnms{Geoff K.}~\snm{Nicholls}\ead[label=e3]{nicholls@stats.ox.ac.uk}\orcid{0000-0002-1595-9041}}
\address[A]{Mathematical Sciences,
University of Southampton\printead[presep={,\ }]{e1}}

\address[B]{Cellmark Forensic Services \printead[presep={,\ }]{e2}}

\address[C]{Department of Statistics,
University of Oxford\printead[presep={,\ }]{e3}}
\end{aug}

\begin{abstract}
Classification of unlabeled data is usually achieved by supervised learning from labeled samples.
Although there exist many sophisticated supervised machine learning methods that can predict the missing labels with a high level of accuracy, they often lack the required transparency in situations where it is important to provide interpretable results and meaningful measures of confidence.
Body fluid classification of forensic casework data is the case in point. 
We develop a new Biclustering Dirichlet Process for Class-assignment with Random Matrices (BDP-CaRMa), with a three-level hierarchy of clustering, and a model-based approach to classification that adapts to block structure in the data matrix.
As the class labels of some observations are missing, the number of rows in the data matrix for each class is unknown. 
BDP-CaRMa handles this and extends existing biclustering methods by simultaneously biclustering multiple matrices each having a randomly variable number of rows.
We demonstrate our method by applying it to the motivating problem, which is the classification of body fluids based on mRNA profiles taken from crime scenes.
The analyses of casework-like data show that our method is interpretable and produces well-calibrated posterior probabilities.
Our model can be more generally applied to other types of data with a similar structure to the forensic data.
\end{abstract}


\end{frontmatter}

\section{Introduction}
Body fluid samples are commonly taken as part of the evidence gathered from a crime scene. However, the fluid-type is often unknown and must be identified. Five body fluid types are of interest here: cervical fluid (CVF), menstrual blood (MTB), saliva (SLV),  blood (BLD), and semen (SMN). 
One forensic method used for body fluid identification is messenger RNA (mRNA) profiling \citep{harbison2016fo}. 
This technique assays samples for the presence of mRNA species (markers) that are characteristic of particular body fluids. 
Markers (i.e., columns) come in fixed groups defined by the fluid-type they target; we call these \emph{marker groups}.
The mRNA signal data is given as a sample/feature matrix, in which the rows are \emph{mRNA profiles} for different specimens, the columns correspond to different mRNA markers, and the matrix entries are binary indicators of marker presence/absence obtained by thresholding a marker amplification response measure \citep{lindenbergh2012mu, akutsu2022pr} described below.
The data matrix shown in Figure~\ref{fig:dataGrid} has 25 fluid-type/marker-group \emph{blocks}.
We call a row of five blocks a \emph{fluid-type matrix}.

Classification of unlabeled fluid-types using binary marker profiles is often straightforward: if we simply choose the fluid-type with the most amplified target-markers (Table ~\ref{tab:typeByMkrCount} in Appendix~\ref{app:CF-simple-majority}), we classify fairly accurately without careful analysis. 
However, the court needs well-calibrated measures of uncertainty for the fluid-type of a given mRNA profile. 
In the following, we refer to ``classifying profiles'' as a shorthand for ``quantifying the uncertainty in the assignment of unlabeled mRNA profiles to fluid-types''.

Likelihood ratios are commonly used in court \citep{morrison21}, and careful statistical modeling is necessary to produce well-calibrated measures. 
Several training data sets of profiles labeled with known true fluid-types are available. 
These are large enough for off-the-shelf machine learning methods, such as random forests and support vector machines (SVM), to be used for classification \citep{tian2020ne, wohlfahrt2023ba}.
Bayesian approaches include \cite{zo2016pr}, using na\"{i}ve Bayes, and the work of \cite{fujimoto2019di}, involving fitting partial least squares-discriminant analysis.
However, these methods do not accommodate heterogeneity in the sample population of mRNA profiles within a fluid-type.
In order to model this heterogeneity, we cluster mRNA profiles within a fluid-type and cluster the mRNA marker signals targeting each particular fluid-type. 
This \emph{biclustering} captures signal patterns in the sample and feature populations.

Matrix biclustering methods, which cluster samples and features simultaneously, are widely used in bioinformatics. 
In our setting and \cite{li2020ba}, samples in a row cluster share similar patterns over features, while feature clusters identify features that share similar patterns over samples within a given row cluster. 
The transpose of this setup, clustering samples within feature clusters, seems more common (see for example, \cite{lee13} and citing literature). 
This conditional biclustering is just one of many bicluster patterns that have found use: we list some of these and their many applications in our literature review.   

In the work we cite, the goal of the inference is often the biclustering itself, though it also supports dimension-reduction for estimation of latent matrix-element parameters \citep{hochreiter2010fa, murua2022Bi, lee13}. 
\emph{Our} main task is to classify new profiles taken from a crime scene, and biclustering is needed to get a model that fits the data and supports profile-classification. 
Biclustering \emph{per se} is of secondary interest.
We develop a Biclustering Dirichlet Process for Class assignment over Random Matrices (BDP-CaRMa) and use it to classify single-source forensic mRNA profiles.

BDP-CaRMa has a three-level hierarchy: the highest level groups profiles (matrix rows) into five fluid-type matrices. 
This grouping is known for the labeled training data. However, fluid-type matrices have a random number of rows, as their row-content depends on the assignment of unlabeled profiles to fluid-types. 
At the second level, all profiles in a fluid-type matrix are partitioned into subtypes; this row-clustering is unknown for all profiles, and so it is a random variable in the posterior. 
Figure~\ref{fig:dataGrid} displays the top two levels of the hierarchy. 
At the third level, markers (i.e., columns) are clustered within row-clusters. 
There is an independent column-clustering of markers within each row-cluster and each marker group. 
A set of cells in the same row- and column cluster is a bicluster.
Figure~\ref{fig:BDP-single-matrix} shows a biclustering of one fluid-type/marker-group block. 
Finally, we have a parameter vector $\theta$ with one component for each bicluster. The data in each cell in a bicluster are iid given the bicluster-parameter.
 


In our forensic setting, each unlabeled profile must be classified one at a time and independently of other unlabeled profiles for legal and ethical reasons. 
Further, data from the crime scene should not influence our model for the training data, so the unlabeled profile should not inform the biclustering of the training data. 
In Bayesian inference the data inform a joint biclustering of all profiles, so in the forensic setting at least, Bayesian inference is ruled out. 
However, \emph{Cut-Models} \citep{liu_modularization_2009,Plummer2015CutsModels} are a form of Generalised Bayesian inference \citep{bissiri_general_2016}, which modulate the flow of information in an analysis. 
In other applications, the opposite is true: joint classification of multiple profiles using Bayesian inference is the belief update with the greatest information gain and would be adopted. 
Our notation in sections \ref{sec:data_obs_model}, \ref{sec:posteriors-all}, \ref{sec:cut-model}, and \ref{sec:mcmc} handles both cases. 
The Bayesian BDP-CaRMa is given in Section~\ref{sec:posteriors-all} and Cut-Model inference in Section~\ref{sec:cut-model}. 

Our models and the computational methods presented here can be applied to other types of forensic data for body fluid classification, such as protein \citep{legg2014di} and microRNA markers \citep{fujimoto2019di,he2020id}.
However, BDP-CaRMa may be useful for supervised class-assignment whenever the class label (our fluid-type) is a categorical response and the features are covariates, or when the class label is a categorical covariate and the features are conditionally independent response values.

\subsection{Our contribution}
We take the (transpose of the) ``NoB-LoC'' biclustering process \citep{lee13} as our starting point for model elaboration. 
Each matrix cell has one or more latent parameters. 
Our biclustering model groups these parameters across cells; all parameters in a bicluster are equal. 
This is not the case in NoB-LoC, so we first modify the distribution of parameters within biclusters and arrive at a model like the BAREB model \citep{li2020ba} for periodontal data. 
Those authors have a two-level biclustering hierarchy and use a multinomial-Dirichlet distribution to define the distribution over clusterings. 
We use the related Dirichlet Process (DP, \cite{ferguson1973ba}) and Multinomial Dirichlet Process (MDP, \cite{Ghosal2017FundamentalsInference}). 
The different approaches are contrasted in Appendix~\ref{app:nested-DP-connection}.


Our inferential goal is to identify the fluid-type of an unlabeled profile. As these profiles move between fluid-types in our Monte-Carlo, the set of rows partitioned by BDP-CaRMa for a given fluid-type is random; it depends on which unlabeled profiles are assigned to that fluid-type. 
Our BDP-CaRMa is therefore a random process partitioning random sets of profiles. This new methodology is needed in applications of biclustering to supervised classification.

Fitting BDP-CaRMa is challenging as each row partition has a ``parameter'' which is itself a random partition, like NoB-LOC and BAREB. 
In NoB-LoC and citing literature, this is handled using carefully adapted reversible jump proposals. 
However, in the setting of our motivating application, we can integrate out all parameters of BDP-CaRMa below the row subtype-clustering exactly; this leaves us with a marginal posterior defined on partitions of the rows of the fluid-type matrices. 
This is all we need, as our inferential goal is to locate unlabeled profiles within fluid-types: we have no need to recover the biclustering itself. 
It also allows straightforward and efficient Markov Chain Monte-Carlo (MCMC) simulation. 

Finally, a user can obtain well-calibrated posterior probabilities for the class-assignment of unlabeled mRNA binary profiles to fluid types: this meets the needs of the court and answers our motivating forensic question. 
We show that the posterior probabilities we estimate are well-calibrated, in the sense that Beta-calibration \citep{kull17} gives recalibrated class probabilities close to the original posterior probabilities.

\subsection{Previous work on forensic body-fluid classification}

We divide work on body-fluid identification into two categories.
The first aims to verify whether a sample belongs to some given fluid-type. For example \cite{akutsu2020de} present a multiplex RT-PCR assay, i.e., a small set of mRNA markers (ESR1, SERPINB13, KLK13, CYP2B7P1, and MUC4) and estimate a likelihood ratio using Bayesian inference.

The second category classifies a sample into one of a small number of candidate fluid-types, as here.
\cite{he2020id} use discriminant analysis with forward stepwise selection to classify a micro RNA profile into one of five fluid-types of interest (peripheral blood, menstrual blood, vaginal secretion, and semen), and \cite{tian2020ne} used a random forest to classify DNA methylation profiles as venous blood, menstrual blood, vaginal fluid, semen, and buccal cells. 
\cite{iacob2019ma} is similar. 
Bacterial community composition data has been used to predict body fluid-types: \citep{wohlfahrt2023ba} use a support vector machine and tree-ensemble methods and reliably distinguish between cervical fluid and menstrual blood, a challenge for mRNA molecular data. 
While these authors all demonstrate accurate prediction with their model, they do not attempt to quantify uncertainty in-class assignments.

Methods suitable for samples, which contain a mixture of different fluid-types, have also been developed \citep{akutsu2022pr}. 
Among these, \cite{Ypma2021} analyzes mixed-fluid data using neural nets and a random forest in a frequentist setting. 
They use a form of Platt scaling \citep{platt2000} resembling Beta-scaling to calibrate likelihood ratios. 
Further details on data types used for body fluid identification are discussed in \cite{sijen2015mo}.

\subsection{Previous work on biclustering}\label{sec:literature-review}

The ``NoB-LoC method'' \citep{lee13} is a model for biclustering with a nested structure. 
The authors apply it to protein expression level data from breast cancer patients.
The method identifies subgroups of proteins (columns there) and then clusters the samples (rows in that setting) within each protein subgroup to give biclusters: within each bicluster, $\theta$-parameters are shared across samples but not across proteins.
\cite{zuanetti2018cl} use a Nested Dirichlet Process (NDP, \cite{Rodriguez2012TheProcess}) to identify clusters of DNA mismatch repair genes based on their gene-gene interactions and those of microRNA based on binding strength across different genes.
These papers work in the ``marignalized'' setting, where the partitions are explicit but the infinite-dimensional Dirichlet process itself is integrated out.

\cite{xu2013no} and \cite{zanini2023dependent} build on \cite{lee13}: 
\cite{xu2013no} gives a nonparametric Bayesian local clustering Poisson model (NoB-LCP) to infer the 
biclustering of histone modifications and genomic locations; \cite{zanini2023dependent} extends NoB-LoC to handle protein expression data from lung cancer patients to identify clusters of proteins and the clusters of patients and cell lines nested therein. 
Like NoB-LoC, the parameters $\theta$ are independent across columns within a bicluster. 

\cite{li2020ba} work in a similar setting to \cite{lee13}. 
However, the entries in the matrix that they bicluster are not response-values but covariate-values in a linear model with an independent response for each matrix row. 
The $\theta$-parameters in each cell are effects, which are equal across cells in a bicluster. 
\cite{yan2022ba} also bicluster a matrix of covariate-effects in a model for a row-response. 
They carry out variable selection within each row partition of the HapMap genomic SNP data, so they select different effects for different clusters of individuals.  
In contrast, each row of our data matrix is a vector of response values with a common covariate (the fluid type for that row).

\cite{li2020ba} take a Multinomial-Dirichlet {\it Distribution} with $J$ categories for both row and nested column partitions. 
This prior, often used for clustering in mixture models, allows empty partition sets; in some parameterizations, the marginal distribution over non-empty sets is the MDP. 
Their ``BAREB-model'' has biclusters distributed like NoB-LoC but is closer to our BDP-CaRMa setup as there is one independent parameter associated with each bicluster (compare Figure~1 in \cite{li2020ba} and Figure~\ref{fig:BDP-single-matrix}). 
They apply their model to the analysis of biomedical dental features measured across tooth-sites and over patients, selecting upper bounds on the number of clusters using the WAIC \citep{watanabe_widely_2012,vehtari2017pr}. 
Like many of the papers developing NoB-LoC for new applications, BAREB-analysis uses Reversible-Jump MCMC to fit the model to data. 
The goal of the inference in BAREB is to estimate effect sizes in a regression model with different parameters for each bicluster, though the biclusters themselves are also of interest. 
The goal of our work is to classify unlabeled profiles, which leads to the main difference between our work and BAREB and NOB-LoC: the number of rows in each matrix we bicluster is random, as the matrix to which an unlabeled profile belongs is unknown. 
We discuss some related branches of the biclustering literature in Appendix~\ref{app:literature_more}.

Many applications of biclustering in bioinformatics use sparse factor-analysis \citep{hochreiter2010fa,moran2021Sp,wang2021em} in which biclustering sets are (possibly overlapping) rectangular subsets of the target sample/feature matrix. 
FABIA \citep{hochreiter2010fa} is widely used for this purpose.
Closely related to independent component analysis (ICA, \cite{Hyvarinen1999su}), it identifies biclusters in gene-expression data using factor-analysis with sparse loadings and factors.
Sparsity is achieved using Laplace priors, while inference is carried out via variational EM to estimate MAP values for factor and loading matrices. 
This gives accurate point estimates but does not feed uncertainty into downstream inference.
In \cite{moran2021Sp}, sparsity is induced using Spike-and-Slab Lasso priors \citep{rovckova2018sp} for factors and loadings.
They identify subtypes of breast cancer from gene expression data and recover major cell types from scRNA expression-level data across cell types, using Bayesian inference to quantify uncertainty. 
\cite{wang2021em} presents a computational framework suitable for fitting very general sparse factorization models, alternating between Empirical-Bayes estimation of prior hyper-parameters and Variational-Bayes (VB) approximation of parameters with the final VB posterior quantifying uncertainty. 

\section{mRNA profile data}\label{sec:mRNA-profile-data}
Our dataset consists of $M = 27$ binary features measured on samples. 
We work with all the data available to us.
These data were provided by forensic scientists who chose markers expected to support fluid-type identification for crime-scene analysis: we do not drop or otherwise select data to simplify the statistical analysis.
Following laboratory processing, the presence/absence of each marker in a sample is visualized in an electropherogram that represents the signal for each marker as a peak height measured in relative fluorescence units. 
The markers listed in Table~\ref{tab:rnaProfile} 
\begin{table*}
\caption{Profile counts and the candidate markers for each body fluid-type in the training and test datasets. 
Two ``housekeeping'' markers are used for quality control. 
They should be amplified in every profile.}
\label{tab:rnaProfile}
\begin{tabular}{@{}|l|c|c|c|@{}}
\hline
Body fluid-type & \multicolumn{2}{c|}{Profile counts} & Candidate Markers \\
\cline{2-3}
 & Train & Test & \\
\hline
Cervical fluid    & 59 & 24 & CYP, HBD1, Lcris, Lgas, MUC4  \\
Menstrual blood    & 31  & 0 & Hs202072, LEFTY2, MMP10, MMP11, MMP7, MSX1, SFRP4  \\
Saliva    & 80  & 10 & HTN3, MUC7, PRB4, SMR3B, STATH  \\
Blood    & 65  & 2 & ALAS2, GlycoA, HBB, PF4, SPTB \\
Semen    & 86  & 10 & MSMB, PRM1, PRM2, SEMG1, TGM4 \\
Housekeeping & N/A & N/A & TEF, UCE\\ 
\hline
\end{tabular}
\end{table*}
were chosen to respond strongly or ``light up'' for a specific fluid-type. 
We have $F = 5$ body fluid-types, coded as cervical fluid (1/CVF), menstrual blood (2/MTB), saliva (3/SLV),  blood (4/BLD) and semen (5/SMN), and five groups of markers, as the markers in each group target a specific fluid-type.
Following standard practice in the field, each raw profile is converted to an $M$-dimensional binary vector (a sample \emph{profile}) using a cut-off threshold: the $i$'th entry in this vector records the amplification status (above or below the threshold) of the $i$'th mRNA marker in the raw profile. 
Details of sample collection and data generation can be found in Appendix~\ref{sec:data-generation}.

 The goal of our analysis is to infer the fluid-type of an unlabeled binary profile. 
 We work with two data sets: a training dataset with 321 profiles and a test dataset with 46. 
 These data are summarised in Table~\ref{tab:rnaProfile}. 
 Roughly speaking the training data are gathered under ``laboratory conditions,'' while the test data are gathered under conditions much closer to casework scenarios.
 We expand on this in Appendix~\ref{sec:data-generation} and return to it in Section~\ref{sec:cut-model-post}.
 All our data are labeled as we know the fluid-type of all the profiles in both datasets. 
 
 We used the training data for model development, using leave-one-out cross-validation (LOOCV) to check performance, where each training profile held-out is treated as unlabeled.
 Subsequently, we tested our methods by treating the profiles in the test dataset as unlabeled. 
 The test data came to us later in our work, and we did not revise our methods after applying them to the training data, so the results we report on the test dataset are ``first shot'' and should be representative of performance on new marker profiles drawn from the same sample population as the test data.
 
\begin{figure}[tp]
        \includegraphics[width=14.5cm]{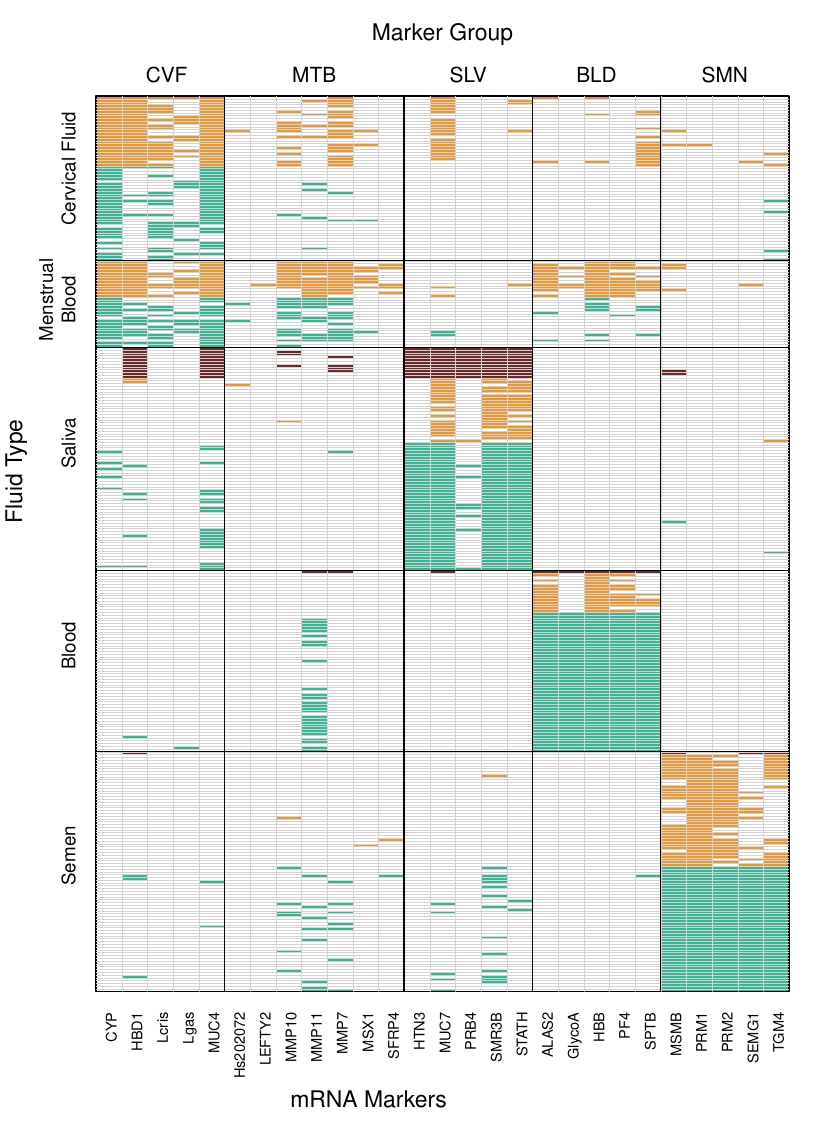}
        \centering
        \caption{Training data, mRNA profiles colored by row-subtype (posterior mode estimated from later analysis) within fluid-type.}
        \label{fig:dataGrid}
\end{figure}

The training data are visualized in Figure~\ref{fig:dataGrid}: each row of Figure~\ref{fig:dataGrid} gives the binary marker profile for a unique sample; each column gives the binary responses of a unique marker.
Rows/profiles are grouped into the five fluid-types of the sample labels, and columns/markers are grouped by the fluid-type they target, giving the 25 blocks.
The binary profiles in the training data show heterogeneity in marker patterns, and in particular, there appear to be subtypes within fluid-types.
We have colored the markers to highlight a possible subtype grouping within each fluid-type 
(using the estimated mode of the posterior in Section~\ref{eq:joint_post_all_FT} to define row-clusters).
This may reflect structure in the population from which the samples are drawn. 
Our analysis must take this unknown subtype structure into account.
For clarity of exposition, in Sections 3--5, we describe our models in a general notation, but very much in the context of our motivating dataset; our method does \emph{not} assume any fixed number of fluid types and markers in the data to be analyzed.

\section{Observation model}\label{sec:data_obs_model}

Suppose we have $T$ profiles from samples with known fluid-type (labeled profiles) and $U$ unlabeled profiles for $N = T + U$ in total. 
Let $\N = \{1, \dots, N\}$, $\M = \{1, \dots, M\}$ and $\F = \{1, \dots, F\}$ denote the profile (row), marker (column) and fluid-type (class label) index-sets. 
Let $\M_g$ be the index-set for markers targeting body-fluid-type $g \in \F$, so that $(\M_1,\dots,\M_F)$ is a partition of $\M$, and $M_g = |\M_g|$ is the number of markers targeting fluid-type $g\in\F$ (for our datasets, seven for MTB and five otherwise).

Denote by $\Y = (y_1, \dots, y_{N})$ a vector of class labels with $y_i\in \F$ for $i\in \N$ giving the fluid-type for the $i$th profile. 
Suppose $y_i$ is known for $i \in \T = \{1,\dots, T\}$ and unknown for $i \in \U = \{T + 1,\dots, T + U\}$. 
Our aim is to infer the missing fluid-type labels 
$y_i,\ i\in \U$ using their binary mRNA marker profiles $x_i = (x_{i,1},\dots,x_{i,M}),\ x_i\in \{0,1\}^M$ as feature vectors and training on the labeled data $(y_i, x_i),\ i\in \T$. 
Let $\X = (x_{i,j})_{i\in \N}^{j\in\M}$ be the $N \times M$ binary marker-response matrix.
In the saturated model
\begin{equation}
x_{i,j} \sim \text{Bernoulli}(\theta_{i,j}),
\label{eq:ampProbDist}
\end{equation}
are conditionally independent, given $\theta=(\theta_{i,j})_{i\in\N}^{j \in \M}$ with $\theta_{i,j} \in \left[0, 1\right]$ for all $i \in \N, j \in \M$.

We reparameterize the missing fluid-type variables $\Y_\U=(y_i)_{i\in \U}$. 
Let 
\begin{equation}\label{eq:V-definition}
    \V_f(\Y_\U)=\{i\in \U: y_i=f\}
\end{equation} 
be the indices of all the unlabeled profiles assigned by some choice of $\Y_\U\in \F^U$ to fluid-type $f\in \F$.
We call the vector, $\V = (\V_1,\dots,\V_F)$,  an ``assignment partition'' of profiles in $\U$ as $\V_f\subseteq \U$ is the set of unlabeled profiles that are assigned to fluid-type $f$. 
The sets in $\V$ are ordered and may be empty: for example, $\V=(\{1,2\},\emptyset,\dots,\emptyset)$ assigns profiles 1 and 2 to CVF while $(\emptyset,\dots,\emptyset,\{1,2\})$ assigns them both to SMN. 
We work with $\V$ and map back to $\Y_\U$ at the end using $y_i(\V)=\{f\in \F: i\in \V_f\}$.

Let $\T_f = \left\{i \in \T: y_i = f\right\}$ be the indices of labeled data in fluid-type $f\in\F$, and let $\N_f = \T_f\cup \V_f$ represent the $N_f = |\N_f|$ profile-indices labeled or assigned by $\V$ to be in fluid-type $f$. 
The sub-matrix $\X_f=[x_{i,j}]_{i\in \N_f}^{j\in \M}$ is the $N_f\times M$ \emph{fluid-type matrix} of observations on fluid-type $f$. 
The full observation model is
\begin{align}
     p(\X|\theta,\V) &=\prod_{f\in\F} p(\X_{f}|\theta,\V) \nonumber \\
     &=\prod_{f\in\F}\prod_{g\in \F}\prod_{i\in \N_f}\prod_{j\in\M_g} \theta_{i,j}^{x_{i,j}}(1-\theta_{i,j})^{1-x_{i,j}}, \label{eq:ampProb}
\end{align}
where conditioning on $\V$ is needed to fix the profiles $i\in\N_f$ assigned to fluid-type $f\in\F$. 

Our ultimate goal is to estimate the assignment partition $\V$ mapping unlabeled profiles to fluid-types using the posterior $\pi(\V|\X)$, integrating over uncertainty in $\theta$ and any other latent variables. 
In the next section, we give a prior model for the parameters $\theta$.



\section{Biclustering prior}
We set out the BDP-CaRMa distribution over parameters and biclusters. 
For comparison with earlier work on NoB-LoC and clarity of exposition, we first specify the BDP (just biclustering, no class assignment) on a single fluid-type/marker group matrix: one of the 25 blocks in Figure~\ref{fig:dataGrid}. 
We then specify the process on a fluid-type matrix and
take a simple product of fluid-type matrix biclusterings. 
Finally, we model uncertainty in the row content of fluid-type matrices to allow unlabeled profiles to migrate across fluid types giving BDP-CaRMa. 
In this section, the assignment of profiles to fluid-types is conditioned on the assignment partition $\V$. 

\subsection{Multinomial Dirichlet Process}\label{sec:mdp}
 The Dirichlet Process (DP) and the associated Chinese Restaurant Process (CRP) are projective models for partitions. 
 However, in order to avoid the tail of small clusters seen in the CRP, we work with a {\it Multinomial} Dirichlet Process (MDP), which sets an upper bound on the maximum number of partition sets.  
The MDP converges (rapidly in our experience) to the DP as this bound is taken to infinity. 
See \cite{Ghosal2017FundamentalsInference} for further discussion.

 Let $\Xi^J_{\A}$ be the set of all partitions of $\A = \{1, \dots, A\},\ A\ge 1$ into at most $J\ge 1$ sets. 
 Suppose $Q\sim \MDP(\alpha, H; J)$ is a generic MDP with size parameter $\alpha >0$, base distribution $H$, and the maximum number of clusters $J$. 
 If $\psi_i \stackrel{\text{iid}}{\sim} Q$ for $i \in \{1, \dots, A\}$, then $\psi = \{\psi_1, ..., \psi_A\}$ is equivalently given by taking a random partition $P \sim P_{\alpha,J}(\cdot)$ from the distribution over $\Xi^J_\A$ given in \eqref{eq:crp_part_prob} below, simulating parameters $\psi^*_k\sim H$ independently for all $k \in \{1, \dots, K\}$ and setting $\psi_i = \psi^*_{k_i}$ with $k_i=\{k: i\in P_k\}$, as in the DP.
 
 Let $P\sim \MCRP(\alpha,J;\A)$ be the multinomial CRP associated with $\MDP(\alpha,H;J)$. 
 The probability for partition $P\in \Xi^J_\A$ is
\begin{equation}\label{eq:crp_part_prob}
      P_{\alpha,J}(P)=\frac{\Gamma(\alpha)}{\Gamma(\alpha/J)^{K}}\frac{J!}{(J-K)!}\frac{\prod_{k=1}^{K}\Gamma(\alpha/J+|P_k|)}{\Gamma(\alpha+A)}\;\mathbb{I}_{1\le K\le J},
\end{equation}
where $|P_k|$ is the number of elements in cluster $k\in \{1,\dots,K\}$, and the indicator function $\mathbb{I}_{1\le K\le J}$ evaluates to 1 if $K \in \{1,...,J\}$ and 0 otherwise. 
This distribution can be simulated by an arrival-process like the CRP, so it is exchangeable and projective and can be simulated using Gibbs sampling. 
However, we use Metropolis-Hastings updates, so \eqref{eq:crp_part_prob} is sufficient.

The MDP is not taken for computational convenience here but on subjective grounds following prior elicitation. 
Replacing the MDP with the DP leads to some simplification; all the marginals derived below are still tractable. 
In later sections, we perform model comparison with DP-like priors (large $J$) and find our model is strongly favored.

\subsection{Biclustering a single matrix}\label{sec:bicluster-single-matrix}

The notation in this sub-section is set up for biclustering a single fluid-type/marker group matrix for clarity of exposition. 
It holds for this section and Appendix~\ref{app:nested-DP-connection} only. 
In later sections, we expand the model to handle a grid of sub-matrices. 

The value $x_{i,j}\in\{0,1\}$ in cell $(i,j)\in \N\times\M$ has an observation model $x_{i,j}\sim p(\cdot|\theta_{i,j})$ with parameter $\theta_{i,j}\in [0,1]$, conditionally independent within each cell. 
Insight gathered from the training data informs the prior for the $\theta$-parameters. 
The rows of the training data in Figure~\ref{fig:dataGrid} have been sorted and colored to highlight a possible row-clustering of profiles suggesting a group structure in the population of sample profiles. 
We call row-clusters within a fluid-type $f \in \F$ the subtypes of $f$: rows within a subtype have similar marker profile patterns, like a barcode. 
However, \emph{the subtype grouping is unknown: the coloring in Figure~\ref{fig:dataGrid} only illustrates one of many plausible subtype groupings}. 
Column clustering is applied separately to each marker group as markers in the same group are clearly correlated. 
In this case, columns in the same group have similar \emph{proportions} of amplified markers. 
As we move from one (row) subtype to another, the column clustering changes. 
This suggests a nested clustering like BAREB \citep{li2020ba}: cluster rows within fluid-type and columns within row subtypes; a bicluster is a group of matrix cells $(i,j)$ in the same column-cluster of a row subtype; all cells $(i,j)$ in a bicluster get the same $\theta_{i,j}$-value. 
%
%

\begin{figure}
    \centering
    \includegraphics[width=3in]{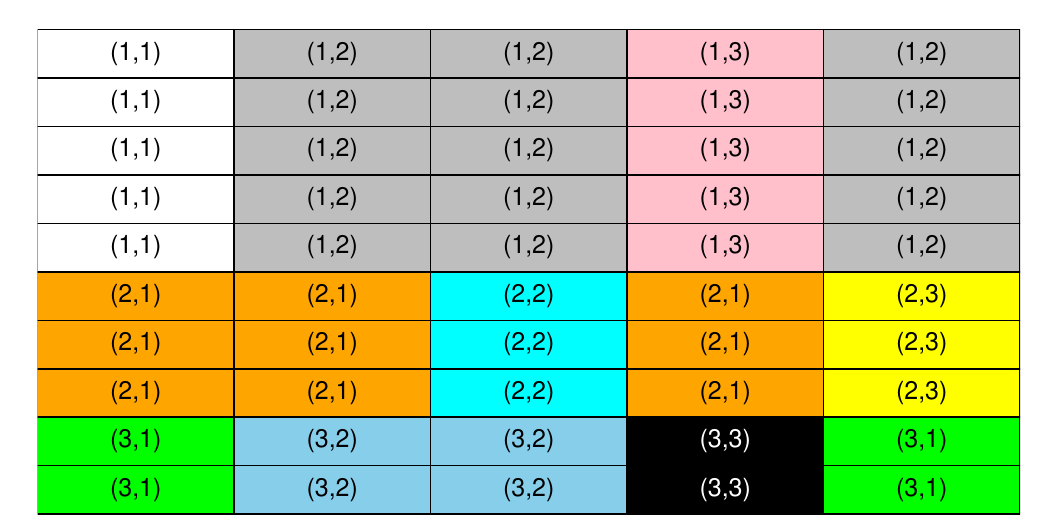}
    \caption{A possible biclustering of a single matrix. Cells with the same color are in the same bicluster. The text within cells gives the $(r,s)$ bicluster label, with $r\in\{1,...,K\}$ giving the row cluster label in the row partition $R = (R_1, ..., R_K)$ of $\{1,...,N\}$, and $s\in\{1,...,K_r\}$ giving the column cluster label in the partition $S_r = (S_{r, 1}, ..., S_{r, K_r})$ of $\{1,...,M\}$.}
    \label{fig:BDP-single-matrix}
\end{figure}

Motivated by this visualization of the data, we use an MDP to partition the $N$ rows into $K$ sets $R = (R_1, ..., R_K)$ with $R_k \subset \{1, ..., N\}$ for all $k \in  \{1, ..., K\}$.
Then, we partition the $M$ columns within each row-cluster $r \in\{1, ..., K\}$ into $K_r$ sets $S_r =(S_{r,1}, ..., S_{r, K_r})$ with $S_{r,s} \subset \{1, ..., M\}$ for each $s \in \{1, ..., M\}$, taking $S_r$ independent of $S_{r'}$ for $r\ne r'$. 
This process partitions the matrix entries $\{1,...,N\}\times\{1,...,M\}$ into sets or ``biclusters'' $C_{r, s} = R_r \times S_{r,s}$.
A possible biclustering of a single $10 \times 5$ matrix is shown in Figure~\ref{fig:BDP-single-matrix}. 
This corresponds to one of the 25 blocks in Figure~\ref{fig:dataGrid} (with fewer rows for ease of viewing).

We specify the biclustering prior in terms of the MDP for partitions, so the priors $\pi_R(R)$ and $\pi_S(S_r),\ r\in\{1,..., K\}$ are given in \eqref{eq:crp_part_prob}. 
We set $\theta_{i,j} = \theta^*_{r,s}$ for all $(i,j)\in C_{r,s}$, with $\theta^*_{r,s}\sim h(\cdot)$ and $h$ the base prior in the overall nested DP. 
The Biclustering Dirichlet Process (BDP) posterior for a single fluid-type/marker group sub-matrix can be written
\begin{equation}\label{eq:post-single-f-single-g}
\pi_{\text{BDP}}(\theta^*,R,S|\X)\propto\pi_R(R)\prod_{r=1}^K \pi_S(S_r)\prod_{s=1}^{K_r} h(\theta^*_{r,s})\prod_{(i,j)\in C_{r,c}} p(x_{i,j}|\theta^*_{r,s}).
\end{equation}
For further details of the relations between the BDP, NoB-LoC, and the NDP, in terms of DP, see realizations, see Appendix~\ref{app:nested-DP-connection}.

\subsection{The Biclustering Dirichlet Process}\label{sec:bicluster}

The BDP defined in Section~\ref{sec:bicluster-single-matrix} was restricted to a single fluid-type/marker group sub-matrix. 
We now extend this to the setting with multiple fluid-types and marker groups. 
Here, all partitions in $R$ and $S$ pick up an additional fluid-type subscript $f \in \F$, and the partitions in $S$ have an extra marker-group subscript, $g\in \F$. 
Examples of the objects defined below are given in Appendix~\ref{app:notation_example} and Figure~\ref{fig:simdat}. 

\subsubsection{Row clusters} \label{sec:row-clusters-prior}
For each $f\in \F$, $R_f = \{R_{f(1)}, \dots, R_{f(K_f)}\}$ is a partition $R_f\in \Xi^{\, J_f}_{\N_f}$ of the row indexes $\N_f = \T_f\cup \V_f$ of $\X$, labeled or assigned by $\V$ to fluid-type $f$. Here $R_f$ has $K_f \in \{1, \dots, J_f\}$ disjoint subsets $R_{f(k)}\subset \N_f,\ k=1,\dots,K_f$. 
We take $R_f \sim \MCRP(\alpha_f, J_f; \N_f)$, so the prior probability distribution $\pi_R$ for $R_f$ as 
\begin{equation}\label{eq:row-cluster-prior}
\pi_R(R_f)=P_{\alpha_f,J_f}(R_f),
\end{equation}
where $P_{\alpha_f,J_f}(R_f)$ is given in \eqref{eq:crp_part_prob}. 
The sub-matrix
\[
\X_{f(k)}=[x_{i,j}]_{i\in R_{f(k)}}^{j\in \M}
\]
represent the data in the ``band" of rows in subtype $R_f$. 

\subsubsection{Column clusters within row clusters} \label{sec:column-clusters-prior}
We partition the columns $\M_g$ of $\X_{f(k)}$ within each marker group $g\in\F$ independently within each row subtype $f(k)$. 
This partition is informed by the data
\[
\X_{f(k),g}=[x_{i,j}]_{i\in R_{f(k)}}^{j\in \M_g},
\]
in $\X$, where the columns in $\M_g$ intersect the rows in $R_{f(k)}$. 
For marker group $g \in \F$, let $L_g \ge 1$ be the maximum number of column clusters permitted, and $\Xi^{L_g}_{\M_g}$ be the set of all column partitions with at most $L_g$ clusters.
We define a column partition $S_{f(k),g} \in \Xi^{L_g}_{\M_g}$ of the columns in marker group $g$ (within row subtype $k$ of fluid-type $f$) as a random partition of $\M_g$ into $K_{f(k),g}\le L_g$ disjoint sets, so that
\[
S_{f(k),g} = \{S_{f(k),g(1)}, \dots, S_{f(k),g(K_{f(k),g})}\}.
\]
``Column subtype'' $l\in\{1,\dots,K_{f(k),g}\}$ of row subtype $f(k)$ is indexed using the notation $(f(k),g(l))$, and this is the basic bicluster label.

Our prior model for column-partitions of a marker type $g\in \F$ within a row subtype $f(k)$ is an MCRP with size parameter $\beta_g > 0$ so that $S_{f(k),g}\sim \MCRP(\beta_g,L_g;\M_g)$, and
\begin{equation}\label{eq:column-cluster-prior}
\pi_S(S_{f(k),g}) = P_{\beta_g,L_g}(S_{f(k), g}),
\end{equation}
where again $P_{\beta_g,L_g}(S_{f(k),g})$ is given by substituting its parameters and argument into \eqref{eq:crp_part_prob}.  

\subsubsection{Biclusters} 
Let $R = (R_f)_{f\in \F}$ be the list of row partitions for the different fluid-types (a ``joint partition''). 
For $f\in \F$ and $k \in \{1,\dots,K_f\}$, let $S_{f(k)}=(S_{f(k),g})_{g\in F}$ be the column partitions within row subtype $f(k)$, we have $S_f=(S_{f(k)})_{k=1}^{K_f}$ and $S=(S_f)_{f\in \F}$.

Given $R,S$, and for all $f,g\in \F$, $k\in \{1,..,K_{f}\}$ and $l\in \{1,\dots,K_{f(k),g}\}$, we define
\[
C_{f(k),g(l)} = \{ (i,j)\in \N\times \M : i\in R_{f(k)}, j\in S_{f(k),g(l)}\},
\]
to be the set of matrix cells in bicluster $(f(k),g(l))$. Let
\[
\C_{f(k),g} = \bigcup_{l=1}^{K_{f(k),g}} \{(f(k),g(l))\},
\] 
be the set of biclusters partitioning the matrix cells in $R_{f(k)}\times \M_g$ and let
$\C_{f,g} = \cup_{k=1}^{K_f}\, \C_{f(k),g}$ be the set of biclusters partitioning the fluid/marker block $\N_f\times \M_g$. 
Finally,
\[
\C_f=\bigcup_{g\in \F} \C_{f,g}
\]
is the set of biclusters partitioning the cells $\N_f \times \M$ in fluid-type $f$ and $\C=\cup_{f\in \F} \C_f$ is the set of biclusters partitioning $\N\times\M$. 

\subsubsection{Bicluster parameters} \label{sec:theta-star-theta-prior}
Each bicluster $(f(k),g(l))\in\C$ has an observation model parameter $\theta^*_{f(k),g(l)}$ for the data in that bicluster. We now elicit their priors. 
 
The mRNA markers $j \in \M_g$ for marker type $g\in \F$ were chosen for measurement by the forensic scientists as likely to amplify for fluid samples of type $f = g$, so we expect $\theta^*_{f(k),g(l)}$ to be large when $f = g$. 
However, some markers are known to be less reliable indicators and fail to be amplified for ``their" target fluid-type. 
This shows up as empty cells in diagonal blocks in the data in Figure~\ref{fig:dataGrid}. 
Also, some off-diagonal blocks show a strong response for an off-target fluid-type (for example, in the MTB profiles, the CVF and BLD markers are frequently amplified).
We take as our prior
\[\theta^*_{f(k),g(l)} \sim \mbox{Beta}(a_{f,g},b_{f,g}),\]  
with different hyper-parameters $a_{f,g} > 0$ and $b_{f,g} > 0$ in each block. 
Fixed values for the prior hyper-parameters $a_{f,g}$ and $b_{f,g}$ are elicited using the training data in Appendix~\ref{sec:fixing-a-and-b}.
We experimented with treating the $a$'s and $b$'s as random variables in the posterior. This is feasible, but slowed things down without any obvious gain.
Let
\[
\theta^*_f = (\theta^*_{f(k),g(l)})_{f(k),g(l)\in\C_{f}}
\] 
be the set of all base parameters for partitions in fluid-type $f$, and let $\theta^*=(\theta^*_f)_{f\in \F}$ be the set of all base parameters.
The base parameters $\theta^*$ are mapped to the original parameters by
\begin{equation}\label{eq:theta_map}
\theta_{i,j}=\theta_{i,j}(\theta^*,R,S),
\end{equation}
where, for $(i,j)\in \N\times\M$, 
$\theta_{i,j}(\theta^*,R,S) = \theta^*_{f(k), g(l)}$, for all $(i,j)\in C(f(k), g(l))$.

\section{The Biclustering Posterior Distribution} \label{sec:posteriors-all}

This section sets out the posterior for the partition $\V$ assigning unlabeled profiles to fluid-types. 

\subsection{Likelihood}
\label{sec:lkd-single-fluid-type}

The parameters in the saturated observation model in \eqref{eq:ampProbDist} are now given by \eqref{eq:theta_map}. The observation model for data $\X_{f(k),g(l)} = [x_{i,j}]_{(i,j)\in C_{f(k),g(l)}}$ in a single bicluster is 
\[
x_{i,j}\sim \mbox{Bernoulli}(\theta^*_{f(k),g(l)}),
\]
jointly independent for $(i,j)\in C_{f(k),g(l)}$. The likelihood for data in a given fluid-type $f\in \F$, row subtype $k\in\{1,\dots, K_f\}$, marker type $g\in \F$ and marker subtype $l \in \{1, \dots, K_{f(k),g}\}$ is
\begin{align}
    p(\X_{f(k),g(l)}|\theta^*_{f(k),g(l)},  R_{f(k)}, S_{f(k),g(l)})  
    &= \prod_{(i,j)\in C_{f(k),g(l)}} (\theta^*_{f(k),g(l)})^{x_{i,j}}(1-\theta^*_{f(k),g(l)})^{1-x_{i,j}} \nonumber
    \\[0.1in]
     &= 
     (\theta^*_{f(k),g(l)})^{s_{f(k),g(l)}}(1-\theta^*_{f(k),g(l)})^{c_{f(k),g(l)} -s_{f(k),g(l)}},\label{eq:base-lkd-cs}
\end{align}
where 
$
c_{f(k),g(l)} = |C_{f(k),g(l)}| 
$
is the number of cells in bicluster $(f(k),g(l))$ and
\[
s_{f(k),g(l)}=\sum_{(i,j)\in C_{f(k),g(l)}} x_{i,j}
\] 
is the number of $1$'s in $\X_{f(k),g(l)}$.

\subsection{Biclustering with a fixed assignment of unlabeled profiles to fluid-types}



Suppose the assignment partition $\V$ of unlabeled profiles $i\in \U$ to fluid-types is fixed, so for $f\in\F$ the set of elements  $\N_f$ partitioned by $R_f$, is fixed.
The posterior distribution of the parameters $\theta^*_f, R_f, S_f$ associated with fluid-type $f$, given all data for fluid-type $f$, is
\begin{align}
    \pi(\theta^*_f, R_f, S_f|\X_f) &\propto  \pi_R(R_f)\prod_{k=1}^{K_f}\prod_{g\in \F}\pi_S(S_{f(k),g})\nonumber\\
    &\qquad\times\quad \prod_{l=1}^{K_{f(k),g}} h_{f,g}(\theta^*_{f(k),g(l)})p(\X_{f(k),g(l)}|\theta^*_{f(k),g(l)}, R_{f(k)}, S_{f(k),g(l)}),
\end{align}
where $\pi_R(\cdot)$ and $\pi_S(\cdot)$ are given in Equations~\ref{eq:row-cluster-prior} and \ref{eq:column-cluster-prior}, and $h_{f,g}$ is the density of a Beta$(a_{f,g},b_{f,g})$ distribution for fixed values of $a_{f,g}$ and $b_{f,g}$ elicited in Appendix~\ref{sec:fixing-a-and-b}.
The $\theta^*_{f(k),g(l)}$-parameters can be integrated out, as their prior $h(\cdot)$ is conjugate, giving 
\begin{align*}
    \pi(R_f, S_{f}| X_f) 
    &\propto \pi_R(R_f)\prod_{k=1}^{K_f}\prod_{g\in \F}\pi_S(S_{f(k),g}) \, p(\X_{f(k),g}|R_{f(k)},S_{f(k),g}),
\end{align*}
where
\begin{align}
   p(\X_{f(k),g}|R_{f(k)},S_{f(k),g})&=\prod_{l=1}^{K_{f(k),g}} 
   \frac{\B(a_{f,g}+s_{f(k),g(l)},b_{f,g}+c_{f(k),g(l)}-s_{f(k),g(l)})}{\B(a_{f,g},b_{f,g})},
   \label{eq:marg_lkd_rowcol_clusters}
   \end{align}
   is given in terms of Beta-functions, $\B(\cdot,\cdot)$.
We further marginalize over partitions of the marker groups: using our data as an example, we have $M_g = 5$ (for four marker groups $g=1, 3, 4, 5$) and $M_g = 7$ (for $g = 2$); summation over $S_{f(k),g}\in \Xi^{L_g}_{\M_g}$ is tractable because the number of partitions of five objects is 52 and 877 for seven. 
We have
\begin{align}
    \pi_R(R_f | \X_f) &\propto \pi_R(R_f)\,p(\X_f|R_f),
    \label{eq:posterior-basic-marginal-R}
    \end{align}
a posterior over $\Xi_{\N_f}^{J_f}$, in which $p(\X_f|R_f)=\prod_{k=1}^{K_f} p(\X_{f(k)}|R_{f(k)})$ with
\begin{align}
    p(\X_{f(k)}|R_{f(k)})&=\prod_{g\in \F} 
    \sum_{S_{f(k),g}\in \Xi^{L_g}_{\M_g}}\pi_S(S_{f(k),g}) \, p(\X_{f(k),g}|R_{f(k)},S_{f(k),g})
    \label{eq:Xfk_given_Rfk}
\end{align}
and $p(\X_{f(k),g}|R_f,S_{f(k),g})$ given in \eqref{eq:marg_lkd_rowcol_clusters}. 
These marginalizations simplify our MCMC as we do not need to propose new column partitions $S_{f(K_f+1),g},\ g\in \F$ when adding row-clusters. 

If the assignment partition $\V$ is fixed, the BDP posterior distribution for the joint partition $R=(R_1,\dots,R_F)$ is simply the product,
\begin{align}
\pi_R(R|\X) 
&\propto \prod_{f\in \F} \pi_R(R_f)\,p(\X_f|R_f),\quad R\in \Xi^{\, *}_\N,
\label{eq:joint_post_all_FT}
\end{align}
with $\pi_R(R_f|\X_f)$ given in \eqref{eq:posterior-basic-marginal-R} and $\Xi^{\, *}_\N$ the product space of joint partitions
\[
\Xi^{\, *}_\N= \Xi^{\, J_1}_{\N_1}\times \Xi^{\, J_2}_{\N_2} \times \dots \times \Xi^{\, J_F}_{\N_F}.
\]
No parameter is shared across fluid-types in this simple product of posteriors.

\subsection{Biclustering with an unknown assignment of unlabeled profiles to fluid-types}\label{sec:post_missing_fluid_types}
 
We now give the posterior for the assignment partition $\V$, taking us from the BDP to BDP-CaRMa. 
Recall the notation of Section~\ref{sec:data_obs_model}: $\T$ and $\U$ are index sets for labeled and unlabeled profiles; $\V=(\V_1,\dots,\V_F)$ is a partition of $\U$ assigning unlabeled profiles to fluid-types. 
The set of labeled profiles in fluid-type $f$ is $\T_f=\{i\in\T: y_i=f\}$, and $\N_f=V_f\cup \T_f$ is the set of profiles labeled or assigned to fluid-type $f$. 
Let
\[
\Xi^{\, *}_{\V,\T}=\Xi^{\, J_1}_{\V_1\cup \T_1}\times \Xi^{\, J_2}_{\V_2\cup \T_2} \times \dots \times \Xi^{\, J_F}_{\V_F\cup \T_F}
\]
be the set of joint partitions $R=(R_1,\dots,R_F)$ given an assignment $\V$. 
If $\V$ is fixed, then $\N_f=\V_f\cup \T_f$ is fixed, and $\Xi^{\, *}_{\V,\T}$ and $\Xi^{\, *}_{\N}$ above are the same set. 
However, $\V$ is now varying so, recalling $\V=\V(\Y_\U)$ from \eqref{eq:V-definition}, let
\[
\Psi_\U=\bigcup_{y_{\,\U}\in \F^U} \{\,\V(y_{\,\U})\,\}
\]
give the set of all assignment partitions, and finally, take
\[
\Xi^{\, *}_{\U|\T}=\bigcup_{\V\in\Psi_\U} \Xi^{\, *}_{\V,\T},
\]
to be the set of joint partitions, allowing for any assignment of the entries in $\U$ to fluid-types. 

Given a joint partition $R\in \Xi^{\, *}_{\N,\U}$, we know the assignment partition $\V(R)$, since
\begin{equation}\label{eq:v_of_R}
\V_f(R)=\bigcup_{k=1}^{K_f}\, R_{f(k)}\cap \U,\qquad \mbox{for $f\in \F$}
\end{equation}
and this fixes the fluid types $y_{\,\U}$ of unlabeled profiles $i\in \U$ via $y_i(R)=\{f\in \F: i\in \V_f(R)\}$.

We decompose the prior for $R$ into the prior for $R|\V$ and a prior for $\V$, taking \[\pi_\V(\V)=F^{-U},\] so each $y_{\,\U}\in \F^{\,U}$ is equally likely a priori. 
As $\{R\}$ and $\{R,\V(R)\}$ are the same events, 
\begin{align*}
\pi_R(R)&=\pi_{R,\V}(R,\V(R))\\
&=\pi_\V(\V(R)) \prod_{f\in \F}\pi_{R}(R_f|\V_f(R))\\
    &=F^{-U}\prod_{f\in \F} P_{\alpha_f,J_f}(R_f),
\end{align*}
as conditioning on $\V_f(R)$ just fixes the set $\N_f$ that $R_f$ is partitioning. 

The joint BDP-CaRMa posterior with data $\X=(\X_\T,\X_\U)$ and missing fluid-types for profiles in $\U$ is then
\begin{align}\label{eq:missing_type_post}
    \pi_R(R|\X,y_\T) &\propto \pi_R(R|y_\T)\,p(\X|R)  \nonumber\\
    &\propto \prod_{f\in \F} P_{\alpha_f,J_f}(R_f)\,p(\X_f|R_f),\quad R\in \Xi^{\, *}_{\U|\T}
\end{align}
as $F^{-U}$ is constant. The additional conditioning on $y_\T$ in the first line is redundant: it emphasizes that we know $y_\T$ and not $y_\U$, so the space of joint partitions is $R\in \Xi^{\, *}_{\U|\T}$, not $\Xi^{\, *}_{\N}$. Equation~\eqref{eq:missing_type_post} resembles \eqref{eq:joint_post_all_FT}. However, the assignment $\V=\V(R)$ of unlabeled profiles to fluid-types is now random.
Equation~\eqref{eq:missing_type_post} determines $\pi_V(\V|\X,y_\T)$ as $R\sim \pi_R(\cdot|\X,y_\T)$ implies $V(R)\sim \pi_\V(\cdot|\X,y_\T)$.

For example, in Section~\ref{sec:loocv} we estimate the marginal posterior probability for some unlabeled profile $i\in \U$ to have fluid-type $f\in\F$ 
that is,
\[
\Pr(y_i(R)=f|\X,y_\T)=\mbox{E}_{R|\X,y_\T}(\mathbb{I}_{i\in V_f(R)}).
\]
This is estimated using MCMC targeting $\pi_R(R|\X)$.

\subsection{Missing data} \label{sec:missing-data-main} 
If some marker data values $x_{i,j}$ are missing, then they will be omitted from the product in \eqref{eq:base-lkd-cs}. 
In that case, $c_{f(k),g(l)}$ is the number of non-missing cells in $C_{f(k),g(l)}$, and the sum giving $s_{f(k),g(l)}$ runs over non-missing cells. 
The datasets used in this study have no missing values. 
Missing marker data are discussed further in Section~\ref{sec:missing-data-appendix}, where we prove that the posterior distribution of a profile with all-missing entries is uniform over fluid-types. 
We used this property to check our code.

\section{Statistical inference with the Cut-Model}\label{sec:cut-model}
In a forensic setting, data from a crime scene should not influence our model for the training data, so the unlabeled profile should not inform the biclustering of the training data. Bayesian inference makes a joint analysis and will not satisfy this condition. 
Our Bayesian analysis will be useful for general classification tasks, but we have to modify the inference to satisfy this extra condition.
There is a second reason to consider an alternative inference framework. 
As noted in Section~\ref{sec:data_obs_model}, the training and test data are gathered under different conditions (``laboratory'' and ``casework-like''), so we have two data sets with shared parameters and similar but distinct observation models. 

This motivates a robust variant of Bayesian inference called Cut-Model inference \citep{liu_modularization_2009, Plummer2015CutsModels}, a coherent belief update \citep{carmona_semi-modular_2020} in the sense of \cite{bissiri_general_2016}. 
We regard the model that we developed for the labeled training data as potentially misspecified for the unlabeled test data, and we stop the test data from informing the parameters of the labeled data.  

Cut-Model inference is a kind of Bayesian Multiple Imputation with two stages \citep{Plummer2015CutsModels}.
Let $Q\in \Xi^{\, *}_{\T}$ be a joint partition of the labeled training profiles. At the ``imputation stage'' we sample the posterior for $Q|\X_\T$  using only the training data.
At the ``analysis stage'', we sample $R|\X, Q$, the posterior for the joint partition of all profiles, but conditioned on the subtype assignment $Q$, so unlabeled profiles can be added to row-clusters of labeled profiles, or placed in new row clusters, but labeled profiles cannot change row cluster. 
This two-stage setup stops unlabeled profiles from informing the subtype-grouping of the labeled profiles. 

Besides being robust, coherent, and principled, Cut-Model inference is practical \cite{nicholson_interoperability_2022}. 
Because the inference breaks up into stages, which respect the modular labeled/unlabeled structure of the data, it is easy for different teams of researchers to work separately on different parts of the inference.  
At the imputation stage, subtype partitions $Q$ for the labeled data can be sampled in advance. 
This step is time-consuming but only has to be done once and not repeated every time a new collection of unlabeled profiles comes along, as in Bayesian inference. 
At the analysis stage, MCMC simulation of $R|\X,Q$ is fast, converges rapidly, and parallelizes well over different realizations of $Q$ passed through from the imputation stage. 

We find (Section~\ref{sec:results}) that Bayesian and Cut-Model inference give essentially identical results, even when $\U$ has many elements and classification is performed jointly. 
This reflects the fact that there is little evidence for misspecification. 
However, the ``operational'' and ``forensic'' advantages remain, and so we recommend Cut-Model inference for this type of analysis.

\subsection{The Cut-Model Posterior}\label{sec:cut-model-post}

Let $Q = (Q_1, \dots, Q_F)$ give the partitions of the labeled data in each fluid-type (a ``joint partition'' with $Q_f \in \Xi^{\, J_f}_{\T_f}$ and $Q\in \Xi^{\, *}_{\T}$). 
We define
\begin{equation*}
\Xi^{\, J_f}_{\V_f\cup \T_f}(Q_f) = \{P\in \Xi^{\, J_f}_{\V_f\cup \T_f}: P_k\cap \T_f=Q_{f(k)},\ k=1,\dots,K_f\}
\end{equation*}
to be the set of partitions of the labeled and unlabeled data that are consistent with a given partition $Q_f$ of the labeled data in fluid-type $f$,
and let
\begin{equation}\label{eq:cut-R-space-given-QV}
\Xi^{\, *}_{\V,\T}(Q) = \Xi^{\, J_1}_{\V_1\cup \T_1}(Q_1)\times \Xi^{\, J_2}_{\V_2\cup \T_2}(Q_2) \times \dots \times \Xi^{\, J_F}_{\V_F\cup \T_F}(Q_F)
\end{equation}
be the set of joint partitions consistent with $Q$ for fixed assignment partition $\V$. Finally,
\[
\Xi^{\, *}_{\U|\T}(Q) = \bigcup_{\V\in \Psi_{\,\U}} \Xi^{\, *}_{\V,\T}(Q)
\]
denotes the set of all joint partitions of the samples in the full dataset that contain $Q$ as a joint sub-partition. 
If $R\in \Xi^{\, *}_{\U|\T}(Q)$, then $\{R\}$ and $\{Q(R),R,\V(R)\}$ are the same events, since $Q_{f(k)}(R) = R_{f(k)}\cap \T_f$ for $f\in \F$ and $k = \{1,\dots K_f\}$ and $\V=\V(R)$ is given in \eqref{eq:v_of_R}.

Given data $\X=(\X_\T,\X_\U)$, the BDP-CaRMa Cut-Model posterior for $R,Q$ is
\begin{align}\label{eq:missing_type_cut_post}
    \pi_\text{cut}(R|\X,y_\T)&=\pi_\text{cut}(R,Q|\X,y_\T),\quad Q\in \Xi^{\, *}_{\T},\ R\in \Xi^{\, *}_{\U|\T}(Q)\nonumber\\
    &\equiv \pi_R(R|\X,Q)\,\pi_Q(Q|\X_{\T},y_\T).
\end{align}
The factors involved are $\pi_Q(Q|\X_{\T},y_\T)\propto\pi_Q(Q)\,p(\X_\T|Q)$ (conditioning on $y_\T$ means $Q\in \Xi^{\, *}_{\T}$, so this is just Equation~\eqref{eq:joint_post_all_FT} with $R\to Q$ and $\N\to \T$), and
\begin{align}\label{eq:cut_post_R_bit}
    \pi_R(R|\X,Q) &=\pi_R(R|Q)\frac{p(\X|R)}{p(\X|Q)},
    \end{align}
with $p(\X|Q)=\sum_{R\in \Xi^{\, *}_{\U|\T}(Q)} p(\X|R)\pi_R(R|Q)$ an intractable normalizing constant.

Cut-Model inference is not Bayesian inference. 
The relation between the Bayes posterior $\pi_R(R|\X,y_\T)$ in \eqref{eq:missing_type_post} and the Cut posterior in \eqref{eq:missing_type_cut_post} is
\begin{align*}
  \pi_\text{cut}(R|\X,y_\T)&\propto 
  \pi_R(R|\X,y_\T)/{p(\X_\U|\X_\T, Q)},
\end{align*}
so we get the Cut-posterior by weighting the Bayes posterior with the posterior predictive probability for the unlabeled data. 
Returning to Equation~\ref{eq:missing_type_cut_post}, $p(\X|Q)$ in \eqref{eq:cut_post_R_bit}
is an intractable function of $Q$, so we cannot target $\pi_\text{cut}(R|\X,y_\T)$ easily with MCMC.


In order to treat the intractable parameter-dependent constant $p(\X|Q)$, we sample the Cut-posterior in \eqref{eq:missing_type_cut_post} using nested MCMC \citep{Plummer2015CutsModels}: we run MCMC targeting $\pi_Q(Q|\X_\T,y_\T)$ giving samples $\{Q^{(t)}\}_{t=1}^{T_0}$ of partitions of the labeled data and then, for each $t \in \{1, \dots, T_0\}$, simulate a ``side-chain'' $\{R^{(t,t')}\}_{t'=1}^{T_1}$ targeting $\pi_R(R|\X, Q^{(t)})$. 
In the side-chain, the labeled profiles in $Q^{(t)}$ have fixed row-partitions, while the fluid-types and subtypes of the unlabeled profiles are updated. 
The intractable factor $p(\X|Q^{(t)})$ cancels in the Metropolis-Hastings acceptance probability in this second MCMC. 
We run each side-chain to equilibrium keeping only the final state, $R^{(t)} = R^{(t, T_1)}$ for all $t \in \{1,\dots, T_0\}$.
The samples $\{R^{(t)}\}^{T_0}_{t = 1}$ are, asymptotically in $T_0$ and $T_1$, distributed according to $\pi_\text{cut}(R|\X,y_\T)$. 
The downside of nested MCMC is this ''double asymptotic'' convergence to target. 
However, the set $\{Q^{(t)}\}_{t=1}^{T_0}$ can be recycled for many different sets of unlabeled profiles, and the side chains targeting $\pi_R(R|\X, Q^{(t)})$ parallelize perfectly and converge very rapidly.


\section{Markov Chain Monte Carlo methods}\label{sec:mcmc}

We used Metropolis-Hastings Markov chain Monte Carlo (MCMC) to target $\pi_R(R|\X,y_\T)$ and $\pi_\text{cut}(R|\X,y_\T)$.
The MCMC is constructed by specifying our own proposals, each of which operates on one profile at a time to update $R$.
The proposal operation on a labeled profile is restricted to its fluid-type but updates its subtype assignment within the given fluid-type (see Appendix~\ref{sec:MH_fixed_FT}). 
Additional proposals are implemented to update $\V$ (see Appendices~\ref{sec:mcmc-between-types} and \ref{sec:mcmc-gibbs}), and in conjunction with the subtype proposal operation above, they update the classification of unlabeled profiles in $R$. 
Of the updates on $\V$, the Gibbs sampler (in Appendix~\ref{sec:mcmc-gibbs}) selects a new subtype across all subtypes of all fluid-types. 
As this update is rather expensive to compute, we have only used it in the Nested MCMC targeting the Cut-Model for the ``side-chains'' described at the end of Section~\ref{sec:cut-model-post}.
All details of our MCMC, including proposals and acceptance probabilities, are given in Appendix~\ref{sec:mcmc-all} with a summary of typical runtimes, effective sample sizes (ESS), and implementation checks in Appendix~\ref{sec:mcmc-convergence}.

%
%

\section{Method Testing}\label{sec:results}

\subsection{Datasets and Experiments}
 We have two mRNA profile datasets, a training set and a test set, with 321 and 46 profiles respectively.
These data 
summarized in Table \ref{tab:rnaProfile}. 

Our experiments select the model using the training data and perform goodness-of-fit using LOOCV on the training data. We evaluate classification accuracy and calibration using the test data. All experiments are listed in Table~\ref{tab:analysisSummaryESS} in Appendix~\ref{sec:mcmc-convergence} with a brief statement of the purpose of the analysis and ESS values for the associated MCMC.
Prior hyper-parameters $a_{f,g}$ and $b_{f,g}$ are given (Appendix~\ref{sec:fixing-a-and-b}) and fixed across all analyses.

We investigated two classification schemes, 
which we call ``Joint Profile Classification'' (JPC) and ``Single Profile Classification'' (SPC). 
SPC classifies one unlabeled profile in each analysis ($U = 1$), whereas JPC jointly classifies multiple unlabeled profiles at the same time ($U > 1$). 
LOOCV is always SPC, but we experiment with both SPC and JPC analyses of the test data, using Bayesian and Cut-Model inference.
 
\subsection{Model Selection}\label{sec:model-selection}
We first performed model selection on the training data using Bayes Factors and considered 10 models in all. 
We use the posterior in Equation~\ref{eq:joint_post_all_FT}, as unlabeled profiles are absent, and the row-content $\N_f=\T_f$ of each fluid-type is fixed.

We developed a variant of NoB-LoC \citep{lee13}, described in Appendix~\ref{app:lkd-nobloc} and suitable for profile class assignment. 
This new model, which we call NoB-LoC-CaRMa, extends NoB-LoC to model joint biclustering of multiple matrices each having a random number of rows. 
NoB-LoC-CaRMa has a similar relationship to NoB-LoC as our BDP-CaRMa has to BAREB \citep{li2020ba}, adapting the single matrix model for class assignment in each case. 

We took ``null-model'' variants of BDP-CaRMa and NoB-LoC-CaRMa with no biclustering at all ($J_f=L_g=1$), and BDP-CaRMa and NoB-LoC-CaRMa  with $(J_f\in\{1,5,10,15\},L_g=M_g)$.
We denote these Model$(J_f,L_g)$. 
For example, BDP$(1,M_g)$ is BDP-CaRMa with no row-clustering but has column clustering using the MDP with $L_g = M_g,\ g\in \F$.
The models run from MDP-like models with strong bounds on the number of clusters to DP-like models with no effective bounds (bounding $K_f\le 15$ is like removing the bound, as $K_f<10$ for support from the likelihood). 
We focused on $L_g=M_g$ as $S_{f(k),g}$ only partitions a set of $M_g$ elements. 
Bayes factors were estimated using The Candidate's estimator $\hat{B}_{\text{cnd}}$ and a Bridge estimator $\hat B_{\text{brg}}$ (see Appendix~\ref{app:bayes-factors}) with good agreement. 
The Candidate's estimator is exact for comparison of models with $J_f = 1$, as we can integrate $\theta^*$ and $S$. 
Our Bridge estimator is unreliable when the two models have little overlap in their support, so it is only evaluated when feasible (BDP vs. NoB-LoC-CaRMa at $J_5 = 5, 10, 15$).

\begin{table*}[t]
\caption{Bayes factors (log base 10) for model comparison (BDP$(J_f,L_g)$ v. NoB-LoC-CaRMa$(J_f,L_g)$). Positive values indicate evidence for BDP. Asterix-marked values are exact (rounded).}
\label{tab:BF-model-comparison-NBL}
\def\arraystretch{1.75}
\begin{tabular}{|r||c|c|c|c|c|}
 \hline
$(J_f,L_g)$ & $(1,1)$ & $(1,M_g)$ & $(5,M_g)$ & $(10,M_g)$ & $(15,M_g)$ \\ \hline
$\hat{B}_\text{cnd}$ & 195$^*$ & 221$^*$ & 292 & 300 & 314 \\
$\hat{B}_\text{brg}$ &  &  & 285 & 290 & 287 \\
 \hline
\end{tabular}

\vspace*{0.2in}

\caption{Bayes factors (log base 10) for model comparison (BDP$(J_f=5,L_g=M_g)$ v. BDP$(J_f,L_g)$). Positive values indicate evidence for BDP$(J_f=5,L_g=M_g)$.}
\label{tab:BF-model-comparison-BDP}
\def\arraystretch{1.75}
\begin{tabular}{|r||c|c|c|c|c|}
 \hline
$(J_f,L_g)$ & $(1,1)$ & $(1,M_g)$ & $(5,M_g)$ & $(10,M_g)$ & $(15,M_g)$ \\ \hline
$\hat{B}_{\text{cnd}}$ & 167 & 69 & 0 & 2 & 10  \\
 \hline 
\end{tabular}
\end{table*}

Results are summarized in Tables~\ref{tab:BF-model-comparison-NBL} and \ref{tab:BF-model-comparison-BDP}.  
The evidence supporting BDP-CaRMa over NoB-LoC-CaRMa in Table~\ref{tab:BF-model-comparison-NBL} is decisive at each level of model complexity. 
Within BDP-CaRMa models, the evidence supporting BDP$(5,M_g)$ in Table~\ref{tab:BF-model-comparison-BDP} is also decisive. 
The smallest Bayes factor supporting this model is $10^2$ (against BDP$(10, M_g)$).
Some further robustness checks on the choice of $(J_f, L_g)$ are given in Appendix~\ref{app:J-sensitivity-analysis}. 
As BDP$(5,M_g)$ is clearly favored in model selection, we use it in all further analyses. 



\subsection{Leave-one-out cross validation}\label{sec:loocv}

We use LOOCV on the training data to evaluate the goodness-of-fit and classification performance BDP-CaRMa, as the test data have no MTB profiles and just two BLD samples. Also, we preserve the test data for our calibration checks. 
At each fold of LOOCV, the label of one profile $y_i\in\T$ in the training data is withheld so $\U=\{i\},\ U=1$, and Bayesian and Cut-Model inference are used to estimate 
\[p_i(f)=\Pr(y_i(R)=f|\X_\T,y_{\T_{-i}}),\quad i\in\T\] 
as a function of $f\in\F$,  with $\T_{-i}=\T\setminus \{i\}$. 
This 
should be large when $f=y_i$.

\subsubsection{Classifying using the Maximum a Posteriori type} 
\begin{figure}[tp]
        \includegraphics[width=14cm]{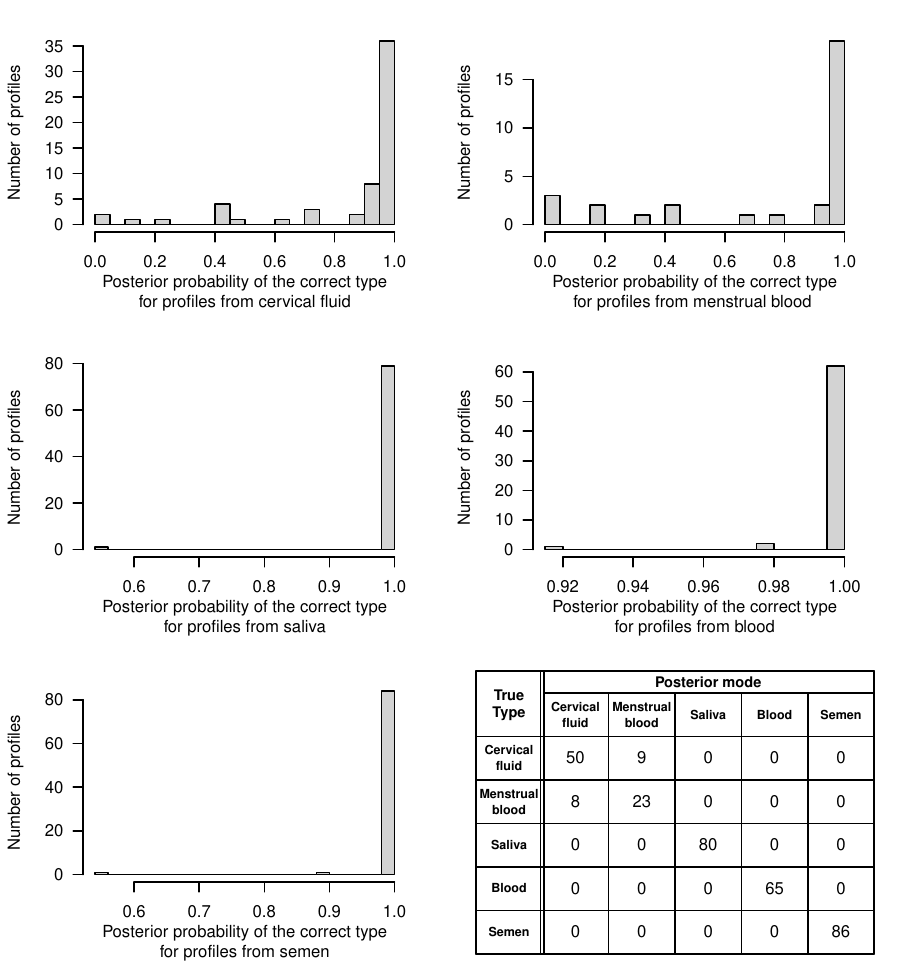}
        \centering
        \caption{(a-e) Posterior probability for the correct type estimated from the LOOCV Cut-Model analysis. (f) Confusion matrix classifying using the posterior mode.}
        \label{fig:cutLOOCV}
\end{figure}

Figure~\ref{fig:cutLOOCV} 
gives results for Cut-Model inference. 
See Figure~\ref{fig:BayesLOOCV} in Appendix~\ref{sec:resultsAppdx} for Bayesian inference, which is near-identical. 
The estimated posterior probability $\widehat p_i(y_i)$ for the correct type is near one for most profiles.
Panels (a)--(e) display the distribution of $\widehat p_i(y_i)$-values over $i\in\T$. 
Panel (f) gives the confusion table obtained from labeling with the mode type $\hat y_i=\arg\max_{f\in\F} \widehat p_i(f)$. 
This is more accurate than na\"{i}ve assignment of types using a majority target-marker rule (Appendix~\ref{app:CF-simple-majority}). 

Some CVF and MTB profiles have low posterior probabilities ($\widehat p_i(y_i)< 0.5$) for the correct type. These fluid types have similar profiles.
For all SLV, BLD and SMN profiles, the posterior mode type is the true type, $\hat y_i=y_i, i\in \N_{3:5}$ and commonly $\widehat p_i(y_i)\approx 1$.

The similarity between Bayesian and Cut-Model inference (Figures~\ref{fig:cutLOOCV} vs. ~\ref{fig:BayesLOOCV}) was expected in a LOOCV/SPC analysis. 
A single profile feeds little information into the fluid subtype partition, so the likelihood for a subtype partition of $\T_{-i}$ is much the same whether the held-out profile $i$ contributes to it (Bayes) or not (Cut). 

\subsubsection{Liklelihood ratios for classification of training data} 
In the analysis in the previous section, the true type may not be the mode for $y_i(R)$ but the evidence for H1: $y_i(R)\ne y_i$ against HO: $y_i(R)=y_i$ 
may be weak. 
The Bayes factor $B^{(i)}_{0,1}=(F-1) \times {p_i(y_i)}/{(1-p_i(y_i))}$ is the prior odds times the posterior odds for these hypotheses. 
Given the focus on likelihood ratios in the courtroom, we ask, how often do we have decisive evidence ($\log_{10} B^{(i)}_{0,1} < -2$, \cite{jeffreys1998th}) against the truth in the LOOCV/SPC analysis? 
Table ~\ref{tab:trainLOOCVCutLR} summarizes Bayes factors computed using Cut-Model inference (results for Bayesian inference are similar).
We find decisive evidence \emph{for} the true type for most $i\in\T$. 
Two out of 321 profiles give ``strong evidence''  \emph{against} their correct type ($-1.5<\log_{10} B^{(i)}_{0,1} < -1$) and one was ``very strongly'' against ($-2<\log_{10} B^{(i)}_{0,1} < -1.5$). These were CVF and MTB profiles.

\begin{table*}[t]
\caption{Do we reject the truth? The number of Bayes factors $\log_{10} B^{(i)}_{0,1},\ i\in\T$ in each interval, for each fluid type, estimated using LOOCV on the training set and Cut-Model inference.}
\label{tab:trainLOOCVCutLR}
\begin{tabular}{ |c||c|c|c|c|c|c|c|}
 \hline
\multirow{2}{*}{True Type}& \multicolumn{7}{c|}{$\log_{10}$ Bayes factor (LR)}\\
 \cline{2-8}
 & $[-2,-1)$ & $[-1,-0.5)$ & $[-0.5,0]$ & $(0,0.5]$ & $(0.5,1]$ & $(1,2]$ & $(2,\infty)$\\
 \hline
CVF & 1 & 1 & 1 & 3 & 7 & 12 & 34\\
MTB & 2 & 1 & 2 & 3 & 1 & 3 & 19\\
SLV & 0 & 0 & 0 & 0 & 1 & 0 & 79\\
BLD & 0 & 0 & 0 & 0 & 0 & 1 & 64\\
SMN & 0 & 0 & 0 & 0 & 1 & 1 & 84\\
 \hline
\end{tabular}

\end{table*}

\subsection{Fluid classification of an independent test set}
\label{sec:classTestSet}

Having selected and tested our BNP-CaRMa model, we now measure its performance as a classifier for the test data.  
We treat the training data as labeled data and the test data as unlabeled data. 
The test data were gathered under conditions designed to mimic casework, and not used in model development, so the likelihood ratios reported here give a better indication of the reliability of the method for classifying mRNA profiles arising in new casework data. 

We compare results from four experiments, pairing Bayesian and Cut-Model inference with SPC (s) and JPC (j) analyses. In the SPC analysis, taking $\U$ to be the unlabeled test data,
\begin{equation}\label{eq:post-prob-single-spc-test}
    p^{(s,m)}_i(f)=\Pr(y_i(R)=f|\X_\T,x_i,y_\T),\quad i\in\U,
\end{equation} 
where $m$ is the posterior for $R$, Bayes ($m=b$) or Cut ($m=c$), while in the JPC analyses
\begin{equation}\label{eq:post-prob-single-jpc-test}
p^{(j,m)}_i(f)=\Pr(y_i(R)=f|\X_\T,\X_\U,y_\T),\quad i\in\U.
\end{equation}

\subsubsection{Comparison across analyses} 
\begin{figure}[tp]
        \includegraphics[width=14.5cm]{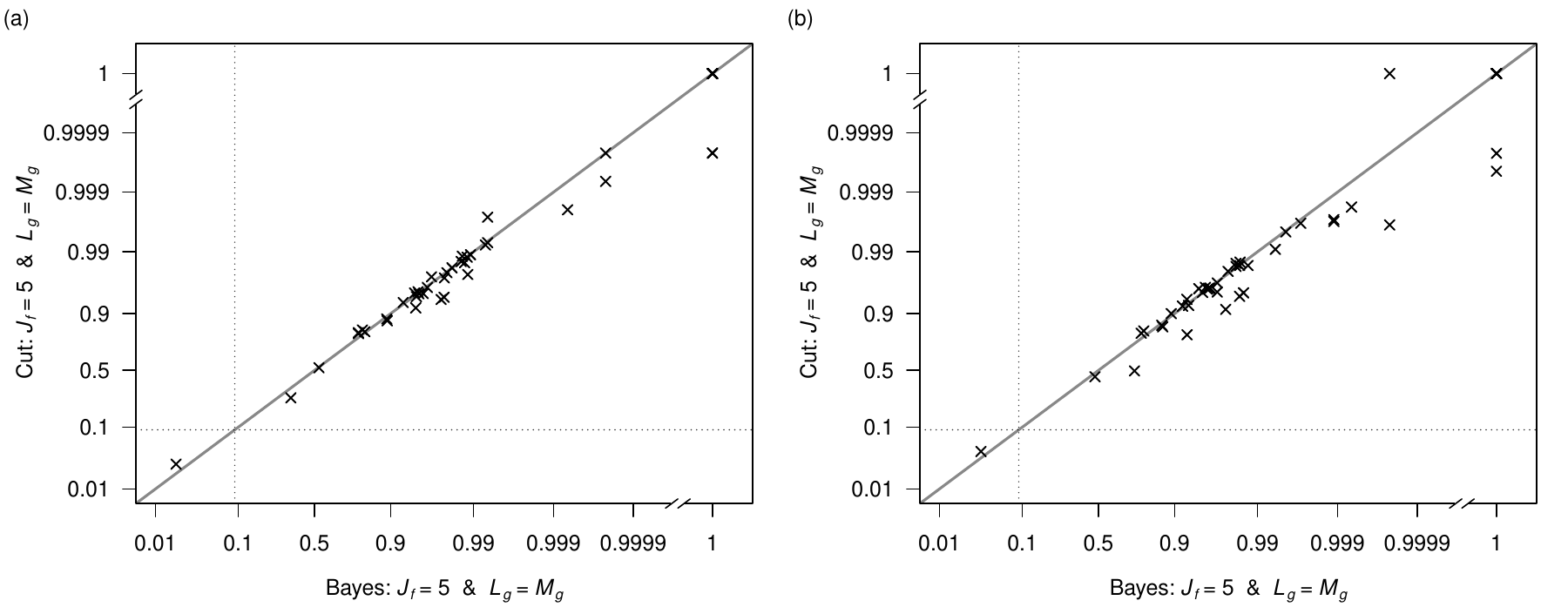}
        \centering
        \caption{Comparison of the posterior probability of the correct types (for mRNA profiles in the test set) between Bayesian- and Cut-Model inference. Panel (a) presents results obtained from analyzing each test mRNA profile one at a time, while panel (b) shows that from analyzing all 46 test mRNA profiles jointly.}
        \label{fig:bayesVsCutCorrectTypePostrDistr}
\end{figure}

Figure ~\ref{fig:bayesVsCutCorrectTypePostrDistr} plots Cut-Model posterior probabilities for true held-out types against Bayes in an SPC analysis (panel (a), $p^{(s,c)}_i(y_i)$ against $p^{(s,b)}_i(y_i)$) and in JPC (panel (b), $p^{(s,c)}_i(y_i)$ against $p^{(s,b)}_i(y_i),\ i\in \U$).
We find that these probabilities are all approximately equal, so the results are robust to the method used for analysis.
This is helpful because the SPC/Cut-Model analysis is very efficient and favored in a forensic setting.
We would expect the JPC/Bayes analysis to have the greatest information gain, but in fact, we lose little in an SPC/Cut-Model analysis.
The similarity of Cut and Bayes analyses also tells us that the sampling distributions of the training and test data are similar.

\subsubsection{Likelihood ratios for classification of test data} Table~\ref{tab:testSingleCutLR} presents Bayes factors $B^{(i)}_{0,1}=4{p^{(s,c)}_i(y_i)}/{(1-p^{(s,c)}_i(y_i))}$ measuring evidence for $y_i(R)=y_i,\ i\in\U$ in the test set using our favored SPC/Cut-Model inference.
For the majority of the mRNA profiles, there is at least ``strong evidence'' ($\log_{10} B^{(i)}_{0,1} > 1$) for the correct body fluid-type over the rest of the four fluid-types.
Of the 46 profiles in the test set, one profile provides ``moderate evidence'' against its true type, a CVF profile favoring MTB.

\begin{table*}[t]
\caption{The number of Bayes factors $\log_{10} B^{(i)}_{0,1},\ i\in\U$ in each interval, for each fluid type in the test set, estimated using SPC/Cut-Model inference.}
\label{tab:testSingleCutLR}
\begin{tabular}{ |c||c|c|c|c|c|c|c|}
 \hline

\multirow{2}{*}{True Type}& \multicolumn{7}{c|}{$\log_{10}$ Bayes factor (LR)}\\
 \cline{2-8}
 & $[-2,-1)$ & $[-1,-0.5)$ & $[-0.5,0]$ & $(0,0.5]$ & $(0.5,1]$ & $(1,2]$ & $(2,\infty)$\\
 \hline
CVF & 0 & 1 & 0 & 0 & 1 & 13 & 9\\
SLV & 0 & 0 & 0 & 0 & 0 & 0 & 10\\
BLD & 0 & 0 & 0 & 0 & 0 & 0 & 2\\
SMN & 0 & 0 & 0 & 1 & 0 & 3 & 6\\
 \hline
\end{tabular}

\end{table*}

\subsection{Calibration}\label{sec:calibrate}

In a forensic setting, we need well-calibrated posterior probabilities for the assignment of fluid-types to unlabeled profiles to ensure that they produce a meaningful measure of the uncertainty in fluid-type classification \citep{dawid1982we, meuwly2017gu,morrison21}.
We use a simple form of Beta-calibration (\citealt{kull17}, Algorithm~1). 
Beta-calibration is Platt-scaling \citep{platt2000} with a careful choice of regression covariates.
 
Suppose the true generative model for a profile $X_i\in \{0,1\}^M$ with true fluid-type $Y_i$ is $Y_i\sim p^*(\cdot)$ and $X_i\sim p^*(\cdot|Y)$, and this holds for $i\in \T\cup\U$ (all training and test data). 
When we observe $X_i = x_i$ for some $i\in\U$, we calculate posterior probabilities $p_i(f;x_i)=\Pr(y_i(R)=f|\X_\T,x_i,y_\T),\ f\in \F$ using our model. 
These are well-calibrated if 
\[
\pi=\Pr(Y_i=f\mid p_i(f;X_i)=\pi).
\]
In analyses of the test data, we have observations $(X_j=x_j,Y_j=y_j),\ j\in\T$ and profiles $x_i,\ i\in\U$ with $Y_i=y_i,\ i\in \U$ held out. 
We estimate  $p_i(f;x_i)$ with $\hat p_i$ using MCMC, and $\hat p_i,\ i\in\U$ are well calibrated posterior probabilities, if 
\begin{equation}\label{eq:recal-good-loocv}
\mathbb{I}_{y_i=f}\sim \mbox{Bernoulli}(\hat p_i),
\end{equation}
for any fixed $f\in\F$. 
We test this using logistic regression. 

We regress $\mathbb{I}_{y_i=f}$ on $z_i=\mbox{logit}(\hat p_i)$ with logistic link and linear predictor $\eta_i=\alpha+\beta z_i$.
In this parameterisation the success probability in the regression $\mathbb{I}_{y_i=f}\sim \mbox{Bernoulli}(p(\eta_i))$ is
\begin{align}\label{eq:beta-cal-p}
    p(\eta_i)=\frac{\hat p_i^\beta}{e^{-\alpha}(1-\hat p_i)^\beta+\hat p_i^\beta}.
\end{align}
This is the simplest of the Beta-calibration maps considered in \cite{kull17}.
If $\alpha=0$ and $\beta=1$ then $p(\eta_i)=\hat p_i$.
Therefore, regressing $\mathbb{I}_{y_i=f}$ on $z_i$ should produce $\hat\alpha\simeq 0$ and $\hat\beta\simeq 1$ if $\hat\pi$ is well calibrated. 
The recalibration map $\mu(\hat p_i)=p(\alpha+\beta\mbox{logit}(\hat p_i))$
is then the identity map. 
Any departure from the identity can be interpreted as an adjustment to $\hat p_i$ required to make $\mu(\hat p_i)$ better match $p^*(y_i|x_i)$.
 
We test calibration for the classification of the test data in all four analyses, SPC/Bayes, SPC/Cut, JPC/Bayes, and JPC/Cut with one calibration regression for each analysis.  
The transformed posterior probabilities in Equations~\eqref{eq:post-prob-single-spc-test} and \eqref{eq:post-prob-single-jpc-test} ($z_i=\mbox{logit}(\hat p^{(s,b)}_i(f))$ etc) are covariates in this regression.
The BLD, SLV, and SMN covariates are linearly separable, and there are no MTB sample profiles in the test data.
However, there are MTB sample profiles in the training data, and the MTB and CVF profiles are often similar, so we can test calibration on the CVF profiles in the test data. 
We re-label profiles as ``CVF'' and ``non-CVF'' and relabel fluid-types $y_i = 1$ (CVF) or $y_i = 0$ (non-CVF) and fix $f = 1$ in \eqref{eq:recal-good-loocv}.

We have a small number of values where $\hat p_i = 0$ or $1$. 
In these cases, we know $0<p_i(f;x_i)<1$, and the error is due to the rounding effect of Monte Carlo with finite sample size. 
Therefore, we take $z_i = f(\hat p_i)$ using a compressed logistic transformation, $f(\hat p_i)=\mbox{logit}((1-2d)\hat\pi+d)$ with $d = 0.0001$. 
We repeated the analysis with the logistic map but dropped profiles with $\hat p_i = 0$ or $1$, and that produced essentially identical results. 
 \begin{table}[t]
     \centering
     \begin{tabular}{r|cccc}
              & SPC/Bayes & JPC/Bayes  &  SPC/Cut & JPC/Cut  \\ \hline
        $\hat\alpha$(s.e.)  & 1.6(9) & 2.0(1) & 1.3(9) &  1.3(9) \\
         $\hat\beta$(s.e.) & 0.9(3)  & 1.1(4) & 0.9(3) & 1.0(4) \\
        $p$-value &  0.15  &  0.11  & 0.23 &  0.27\\
          \hline 
          \multicolumn{5}{c}{} \\
     \end{tabular}
     \caption{Parameter estimates for the calibration map estimated on CVF profiles in the test data. The $p$-values test a null corresponding to the hypothesis that the posterior probabilities are well-calibrated.}
     \label{tab:calibration-tests}
 \end{table}

The fitted recalibration map parameters are given in Table~\ref{tab:calibration-tests}.
We expect these to be correlated as they are computed from the same (test) data. Deviance tests for the null (well-calibrated) model with $(\alpha,\beta) = (0,1)$ against the alternative $(\alpha,\beta)\in R^2$. 
We find no evidence for miss-calibration in any of the analyses $(p\ge 0.05)$. The slope estimates are all close to one. 
Intercepts $\alpha > 0$ indicate that the recalibrated posterior probability is bigger than the uncalibrated posterior probability, uniformly over the latter. 
However, this is not significant. 
There is perhaps a weak case for better calibrated Cut-model inference, as there is less evidence against the null model for the Cut-model analyses.

\section{Concluding remarks}
BDP-CaRMa characterizes patterns in mRNA profiles, offering a flexible and transparent framework for body fluid classification which quantifies uncertainty in the assignment of class labels. 
The well-calibrated probabilistic statements on body fluid classification, which we provide, are important in a forensic setting.
The model has a three-level nested hierarchical structure consisting of the fluid-type, subtype, and marker levels. 
In our classification setting, the assignment of unlabeled profiles to fluid-types is random, so subtypes partition a random set of profiles. 
Related nested biclustering methods \citep{lee13, li2020ba} have two levels of hierarchy and partition fixed sets.

Work by \cite{tian2020ne,wohlfahrt2023ba, Ypma2021}, employing machine learning methods, e.g., random forest, SVM, neural networks, etc., also model heterogeneity in mRNA profiles within a fluid-type. 
Although in some respects simpler than the models given above, our model is well-specified, as evidenced by well-calibrated measures of confidence in assigned class labels. 
Our statistical modeling approach makes explicit any heterogeneity within fluid-types (fluid subtypes). 
Whilst intriguing, these are of secondary interest in our setting. 
However, these structures are likely to be of interest in applications of BDP-CaRMa to classification outside our forensic setting. 

One very helpful feature of our data set and model is that we can integrate out all random variables below the level of fluid subtypes.
The parameter space we actually sample is substantially reduced, facilitating MCMC simulation and making it easier for practitioners to use the tool. 
When the number of columns is large, our MCMC scheme would need to be extended to handle Monte Carlo integration over latent parameters and column clustering via Reversible-Jump MCMC, along the lines of \cite{lee13} and \cite{li2020ba}.
Model selection strongly favors our workhorse BDP$(5, M_g)$ model. 
Sensitivity analysis in Appendix~\ref{app:J-sensitivity-analysis} showed that results are robust to the choice of the maximum number of subtypes when we vary the MDP thresholds $(J_f, L_g)$.

We now make some recommendations on the choice of Bayes or Cut posteriors and the choice of joint or separate analyses. 
Data analysis in a forensic setting is constrained by legal and ethical considerations, so the SPC/Cut-Model analysis is preferred: we would always perform a separate analysis for each profile in $\U$ using Cut-Model inference as it does not allow the casework sample to influence our beliefs about structure in the labeled training data, or interact with each other. 
In applications outside the forensic setting, there will be a loss function, typically some measure of posterior concentration on the unknown true fluid-types. We can minimize the total risk across all profiles jointly or minimize it separately for each unlabeled profile. 
For example, if we have two profiles and ask ``are these the same class?'' then the interaction of the two unlabeled profiles informs their joint class. 
Bayesian inference integrates all the data, so will generally give a lower variance than Cut-Model inference. 
However, Bayesian inference is expected to suffer more bias when the sample populations for labeled and unlabeled data differ. 
In all settings, Cut-model inference has the operational advantages listed in Section~\ref{sec:cut-model-post}. 


In this paper we do not treat sample profiles from non-target materials $f' \notin \F$ or samples which are mixtures of fluid types. 
These profiles are expected to differ strongly from training data.
We can \emph{identify} these ``outlier'' profiles as 
they enter any given fluid-type as a singleton subtype in an SPC analysis. 
A profile with an unusually high posterior probability of being a singleton therefore warrants careful inspection to check data quality and whether it might be none of the candidate fluid-types or has mixed fluid-types. 

In future work, we will extend our model to treat non-body-fluid profiles and mixed-fluid-type profiles. 
We can easily add an extra fluid-type to explicitly accommodate outlier profiles.
A parametric model for mixed-fluid-types also seems in reach of careful statistical modeling and computation, though presents more of a challenge.
In summary, we present a novel classification method using biclustering that provides reliable uncertainty statements and interpretable results on the classification of body fluids for forensic casework, and this provides the foundation for more complex scenarios.

\verb|https://github.com/gknicholls/Forensic-Fluids| gives code and data.

\bibliographystyle{imsart-nameyear} 
\bibliography{references}

\appendix
\newpage
\pagenumbering{arabic}
\setcounter{page}{1}

\hrule
\thispagestyle{empty}
{\begin{center}
    {\bf BICLUSTERING RANDOM MATRIX PARTITIONS WITH AN APPLICATION \\ 
    TO CLASSIFICATION OF FORENSIC BODY FLUIDS}\\[0.1in]

    CHIEH-HSI WU, AMY D. ROEDER AND GEOFF K. NICHOLLS\\[0.15in]

    \Large SUPPLEMENTARY MATERIAL\\[0.15in]
\end{center}}
\hrule
\vspace*{0.15in}

\section{Further literature review}\label{app:literature_more}

In Section~\ref{sec:literature-review}, we cited a selection of biclustering methods most closely related to our own. There are several other Bayesian non-parametric methods for biclustering, taking different biclustering patterns across the target matrix. 
In VariScan \citep{guha2016no}, the column clustering is conditioned on the row clustering, but there is no independent DP within each row-cluster (as we read Equation 2.2 in that work). 
This allows cells in different row clusters to be assigned to the same bicluster. 
Different prior biclustering distributions are appropriate in different applications.
The object of many biclustering analyses is to group exchangeable response values in different matrix cells into biclusters.
Different patterns of biclustering, as in VariScan, NoB-LoC and BNP reflect different prior expectations about the properties of these exchangeable groups.
 
 \cite{guha2016no} elicit a Poisson-Dirichlet-Process (PDP, \cite{perman92}) prior for distributions over row partitions in VariScan. 
 The motivating application in this group of papers is high-throughput gene-expression profile data over patients. 
 We take the simpler Multinomial-Dirichlet-Process (MDP) as our prior for row partitions for similar reasons: it captures our prior expectations about the number and distribution of row- and column-partitions. 
 The upper bound $J$, introduced in Section~\ref{sec:mdp} below, on the number of clusters in our MDP is a prior hyperparameter. 
 If we gave $J$ a Poisson hyperprior, we would have a PDP.

\cite{jha2018no} extends VariScan to mixed data types.
\cite{zhang2019jo} takes the VariScan bicluster structure, in which biclusters can cross row clusters, but replaces the PDP partition prior with an MDD prior of the type used by \cite{li2020ba}, using a grid search to find a good choice of upper bounds on the numbers of row and column clusters.

In other related work, \cite{ren2020bi} cluster time-series data for blood pressure over time and across patients. 
Considering each time-series as a matrix-row, they take a DP over row partitions and then partition the time-series within a row cluster serially, using a change-point process to identify partition boundaries. 

We have focused on Bayesian non-parametric methods, in which the column clustering is nested in, or at least conditional on, the row-clustering. 
However, early parametric biclustering \citep{meeds2007no} took independent Pitman-Yor partition priors on rows and columns, creating a ``plaid'' biclustering in a checkerboard pattern without nesting.
More recently, \cite{murua2022Bi} used a Bayesian plaid model with independent stick-breaking process priors on rows and columns to analyze datasets of histone modifications across genomic locations and gene expression data over time and across genes.

\section{Estimating fluid-type using simple majority}\label{app:CF-simple-majority}

The purpose of this appendix is to show that, if {\it well-calibrated uncertainty measures are not needed}, reasonably {\it accurate} classification of unlabeled profiles may be achieved with very simple methods. 
Table~\ref{tab:typeByMkrCount} presents the confusion table obtained by classifying the fluid type of a profile in the training data by simply taking it to be the fluid type of the marker group with the most amplified markers,
so $\hat y_i=\arg\max_{g\in\F} \sum_{i\in \M_g} x_i$. 
For a row $f\in\F$ in the table, the table entry $N_{f,g}$ in column $g\in\F$ gives $N_{f,g}=\sum_{i\in \T_f} \mathbb{I}_{\hat y_i=g}$.
For example, 17 MTB profiles in the training data were incorrectly classified by this na\"{i}ve method as CVF. 
The approach leads to ties when more than one fluid type have equal-most amplified markers.

\begin{table*}[ht]
\caption{The posterior mode estimated by identifying the marker group with the largest number of markers amplified. Ties are labeled ambiguous.}
\begin{tabular}{ |c||c|c|c|c|c| c|}
 \hline
True  & \multicolumn{6}{c|}{Posterior Mode}\\
 \cline{2-7}
Type  & CVF & MTB & SLV & BLD & SMN & Ambiguous\\
 \hline
Cervical fluid & 55 & 1 & 0 & 0 & 0 & 3\\
Menstrual blood & 17 & 6 & 0 & 0 & 0 & 7\\
Saliva & 0 & 0  & 80 & 0 & 0 & 0 \\
Blood & 0 & 0 & 0 & 65 & 0 & 0 \\
Semen & 0 & 0 & 0 &  0  & 86 & 0 \\
 \hline
\end{tabular}
\label{tab:typeByMkrCount}
\end{table*}

\section{Data Generation}\label{sec:data-generation}
Multiple candidate markers for each body fluid were identified using literature and database searches.
PCR assays were designed for each marker. 
Using known source samples as a template, the markers were then evaluated for amplification in the target body fluid, cross-reactivity with other body fluids, and overall robustness.  
Five to seven markers were chosen for each body fluid and multiplex PCR assays were designed.
In total, the three assays consist of the 27 markers presented in this work and two housekeeping markers. 

The housekeeping-markers'' are used to give an indication of the quality of the sample and as a positive control for the laboratory process. 
They should amplify in every profile.
If they are not detected then the run is discarded. 
They are of no inferential value.

The body fluid samples used for this study were collected from voluntary donors with informed consent.
DNA/RNA co-extraction, DNase treatment, reverse transcription, PCR amplification, PCR purification, and detection for all mRNA analyses were executed based on protocols outlined in \cite{roeder2016}, except an MSMB marker has been used instead KLK3.
For some samples, the mRNA extraction protocol was modified to recover mRNA from EZ1 DNA investigator kit cartridges (Qiagen) post DNA extraction.
Detection of marker amplification was performed by a 3500 Series Genetic Analyzer (ThermoFisher Scientific).
The raw data produces continuous profiles recording the level of fluorescence detected measured in relative fluorescence units (rfu).
The rfu values are an approximation of the quantity (number of PCR amplicons) of each mRNA marker that has been detected.  
The binarised profiles are obtained using marker-specific thresholds for peak detection, which means that a peak height measurement is converted to 1 if it is above the threshold and 0 otherwise.
A minimum peak height threshold of 50 was used in addition to a marker-specific threshold, which was the average plus three standard deviations of the peak heights of that marker, rounded to whole numbers (reverse transcription negative samples were used for these calculations).
These values were set so to minimize the chances of scoring ``background'' as a true body fluid peak.  

All samples used have a single fluid-type.

\section{Connection to Nested Dirichlet Process}
\label{app:nested-DP-connection}




We provide further details setting out the Biclustering Dirichlet Process (BDP) in contrast to the NoB-LoC model \citep{lee13} and the Nested Dirichlet Process (NDP), following \cite{Rodriguez2012TheProcess} and making explicit the DP elements, which were integrated out in our presentation in the text. 
In order to convey the connections of our method with NoB-LoC and NDP, we return to the simple setup of Section~\ref{sec:bicluster-single-matrix} and bicluster a single $N\times M$ matrix as a single unit, dropping the fluid-type and marker-group blocks. 
We further simplify the presentation of BDP by using the DP rather than the MDP as our building block, removing the constraint on the maximum number of clusters, as it can be regarded as a special case. 
The literature cited here typically partitions columns first and then rows within columns, i.e., the row-clustering is nested within column clusters. 
In the following, we use the terminology for our setting, where we cluster columns within row-clusters.

\subsection{NoB-LoC posterior distribution} 
Recall that the BDP priors $\pi_R(R)$ and $\pi_S(S_r)$, $r \in \{1,..., K\}$ are given in \eqref{eq:crp_part_prob}. 
BDP and NoB-LoC take CRP (or MCRP) priors, while BAREB uses a Multinomial-Dirichlet Distribution in a mixture model prior for partitions. 
This has the MDP as its marginal if it is correctly parameterized, and we integrate out empty partitions \citep{Ghosal2017FundamentalsInference}.
NoB-LoC differs from the BDP and BAREB in the way parameters are assigned to cells within biclusters.
In the BDP and BAREB, we simply set $\theta_{i,j} = \theta^*_{r,s}$ for all $(i,j)\in C_{r,s}$, with $\theta^*_{r,s}\sim h(\cdot)$ and $h$ the base prior in the overall nested DP. 
The BDP posterior for a single fluid-type/marker group sub-matrix is given in \eqref{eq:post-single-f-single-g}.
In contrast, NoB-LoC sets $\theta_{i, j} = \theta^*_{i,s}$ for each $j\in S_{r_i,s}$, where $r_i$ is the row-cluster containing row $i$, so it shares $\theta^*_{i,s}$ across all columns $j\in S_{r_i,s}$ but not across different rows $i\in R_r$. 
Its posterior is 
\[
\pi_{\text{NoB-LoC}}(\theta^*,R,S|\X)\propto\pi_R(R)\prod_{r=1}^K \pi_S(S_r)\prod_{s=1}^{K_r} \prod_{i\in R_r} h(\theta^*_{i,s})\prod_{j\in S_{r,s}} p(x_{i,j}|\theta^*_{i,s}).
\]
It may be helpful for understanding the difference to compare this with \eqref{eq:post-single-f-single-g}.

\subsection{Relations in terms of underlying Dirichlet processes}
\label{sec:relDP}
The NDP takes $N$ ``centers'' each with $M$ subjects and clusters the centers, then groups subjects within center clusters. 
The NDP differs from BDP in that there is no notion of ``columns,'' so no specific relation between subject $j$ in center $i$ and subject $j$ in center $i'$. 
In our setting, they are equal, as the same markers $j\in \M$ are measured for each profile $i\in N$. 
If two centers $i,i'$ are in the same cluster in the NDP, then their parameters $(\theta_{i,j})$ and $(\theta_{i',j})$ are sampled from the same DP-realization, so they {\it may be} equal, whereas in the BDP, if two centers are in the same bicluster of the BDP then their parameters {\it must be} equal, so $\theta_{i,j} = \theta_{i',j}$. 

On the other hand, NoB-LoC is a biclustering model for a $N\times M$ matrix $\X=[x_{i,j}]_{i=1,..., N}^{j=1,..., M}$ and hence, does have an explicit notion of ``columns.'' 
NoB-LoC uses a zero enriched P\'{o}lya urn scheme, which permits an extra `inactive' category in the row-partition and the column partitions nested therein. For clarity, we drop that feature in the explanations below (equivalent to sitting $\pi_0 = \pi_1 = 0$ in the original NoB-LoC description).

As in Section~\ref{sec:bicluster-single-matrix}, $R=(R_1,...,R_K)$ is a partition of $\{1,...,N\}$ and for $r \in \{1, \dots, K\}$, we let $S_{k}=(S_{k,1},...,S_{k,K_r})$ be a partition of $\{1,2,...,M\}$. 
We take $\alpha>0,\ \beta>0$ and $H$ a centering or ``base'' distribution on the parameter space of $\theta_{i,j},\ (i,j) \in \N \times \M$.
The generative model for $\theta$ in the NDP is
 \begin{align*}
    Q&\sim \text{DP}(\alpha, \text{DP}(\beta,H)),\\
    G_i&\sim Q,\ i=1,\dots,N,\\
    \theta_{ij}&\sim G_i,\ j=1,\dots,M,
\end{align*}
all jointly independent. 
The NDP can be rewritten in a way that highlights the relation to the BDP and NoB-LoC:
\begin{center}
    \fbox{\begin{minipage}{4.6in}
    \vskip 0.1in
\noindent The generative model for $\theta$ in the NDP is
\begin{align*}
    Q&\sim \text{DP}(\alpha, \text{DP}(\beta,H))\\
    G_i&\sim Q,\ i=1,\dots,N, \mbox{ with $G=(G_1,\dots,G_N)$}
    \intertext{  and $G\to (R,G^*)$ with   $R=(R_1,\dots,R_K)$ and $G^*=(G^*_1,\dots,G^*_K)$}
    \theta_{i,j}&\sim G^*_{k_i},\  \mbox{ for all $(i,j)\in \N\times\M$, where } i\in R_{k_i} \mbox{defines $k_i$}.
\end{align*}
\vskip 0.1in
\end{minipage}
}
\end{center}

We now unpack this generative model, following the presentation in \cite{Rodriguez2012TheProcess}.
If
\[
Q\sim \text{DP}(\alpha, \text{DP}(\beta,H))
\] 
is a realization of a DP with base distribution $\text{DP}(\beta,H)$ then $Q$ can be written
\[
Q=\sum_{t\ge 1} w_t \delta_{Q^*_t},
\]
where $w_t, t\ge 1$ are weights generated by a stick-breaking process, and the atoms $\{Q^*_t\}_{t\ge 1}$ are iid realizations of $\text{DP}(\beta,H)$. 
A set of $N$ independent samples $G_1, ..., G_{N}\sim Q$ is a subset of the atoms of $Q$, and some of the $G_i$'s may be equal. 
We reparameterize $(G_1,...,G_{N})$ as $(R,G^*)$, where $R$ is a partition of $\{1,...,N\}$ and $G^*=(G^*_1,...,G^*_K)$ are the $K$ distinct realisations appearing in $(G_1,...,G_{N})$, with $G_i=G^*_k$ when $i\in R_k$.

Since $G^*_k$ is an atom of $Q$, it is a realization of $\text{DP}(\beta,H)$, so for $k=1,...,K$, 
\[
G^*_k=\sum_{t\ge 1} w_{k,t} \delta_{\psi^*_{k,t}}
\]
where $w_{k,t},\ t\ge 1$ are stick breaking weights and $\psi^*_{k,t},\ t\ge 1$ are iid realizations of $H$.  
The NDP now takes $\theta_{i,j}\sim G^*_k$ independently for each $i\in R_k$ and iid for $j=1,...,M$. If $i,i'\in R_k$ are in the same cluster then $\theta_{i,j}$ and $\theta_{i',j}$ have the same sampling distribution $G_i=G_{i'}=G^*_k$, and so we can get $\theta_{i,j}=\theta_{i',j}$ if the same atom in $G^*_k$ is selected. 
However, in the BDP we want $\theta_{i,j}=\theta_{i',j}$ whenever $i,i'\in S_k$, not just with positive probability, so we diverge at this point.

The corresponding generative models for NoB-LoC and the BDP can be written as follows.

\begin{center}
    \fbox{\begin{minipage}{4.6in}
    \vskip 0.1in
\noindent The generative model for $\theta$ in NoB-LoC and the BDP is
\begin{align*}
    Q&\sim \text{DP}(\alpha, \text{DP}(\beta,H))\\
    G_i&\sim Q,\ i=1,\dots,N, \mbox{ with $G=(G_1,\dots,G_N)$}
    \intertext{  and $G\to (R,G^*)$ with   $R=(R_1,\dots,R_K)$ and $G^*=(G^*_1,\dots,G^*_K)$ giving the unique DP-realisations in $G$. Next, }
    \theta_{k,j}&\sim G^*_k,\ k=1,\dots K,\   \mbox{ with $\theta_k=(\theta_{k,1},\dots,\theta_{k,M})$}\intertext{ 
    and $\theta_k\to  (C_{k},\theta^*_k)$ with  $C_k=(C_{k,1},\dots,C_{k,K_k})$ and $\theta^*_k=(\theta^*_{k,1},\dots, \theta^*_{k,K_k})$ giving the unique $\theta$-values in $\theta$. Finally, for $k=1,\dots,K$ and each $i\in R_k$,}
    \mbox{NoB-LoC: }& \mbox{ (1) for each $l=1,\dots,K_k$, draw } \ p_{i,l} \stackrel{\text{iid}}{\sim} H ;\\ &\mbox{ (2) for each $j\in C_{k,l}$, set } \theta_{i,j}=p_{i,l}\\
    \mbox{BDP: }&\mbox{ for each $j\in C_{k,l}$, set } \theta_{i,j} = \theta^*_{k,l}. 
\end{align*}
\vskip 0.1in
\end{minipage}
}
\end{center}


\vskip 0.1in


For each row cluster $k = 1,..., K$ in NoB-LoC and BDP, we sample a single set $\theta_{k}=(\theta_{k,1},..., \theta_{k, M})$ of parameters $\theta^*_{k,j} \stackrel{\text{iid}}{\sim}  G^*_k$, for $j = 1,..., M$ to define the nested column clusters $C_k = (C_{k,1}, ..., C_{k,K_k})$.
We reparameterize $\theta_{k}$ as $(C_{k}, \theta^*_k)$, where $\theta^*_k=(\theta^*_{k,1},..., \theta^*_{k,K_k})$ are the unique entries in $\theta_k$ and $C_{k}=\{C_{k,1},...,C_{k,K_k}\}$ is a partition of $\{1,...,M\}$ satisfying $\theta^*_{k,\ell}= \theta_{k,j}$ when $j\in C_{k,\ell}$.
The following describes a key difference between NoB-LoC and BDP.
For each $(i,j)\in \N\times\M$, in NoB-LoC, we set $\theta_{i,j} = p_{i,l}$ with $p_{i,l} \stackrel{\text{iid}}{\sim} H$, where $i \in R_k$ and $j \in C_{k,l}$, which means that, if $(i, j), (i',j') \in R_k \times C_{k,l}$, $\theta_{i,j} = \theta_{i,j}$ only if $i = i'$, so we share parameters across columns in the same row but not across rows within a bicluster.
On the other hand, in BDP,  $\theta_{i,j} = \theta^*_{k,l}$, so parameters are shared across all rows and columns within a bicluster. 

The induced marginal distributions
\begin{align*}
R&\sim \CRP(\alpha,\N),\\
C_{k}&\sim \CRP(\beta,\M), \quad k=1,...,K,  
\end{align*}
are known,
and $\theta^*_{k,\ell} \stackrel{\text{iid}}{\sim} H$ for each $k$ and $\ell=1,...,K_k$. 
We work directly with these marginal distributions in setting up the BDP, and hence $Q$ and $G^*$ are not needed.

\section{Example of notation for nested clustering}\label{app:notation_example}
\begin{figure}
    \[\text{(a)}\quad \includegraphics[width=2.65in]{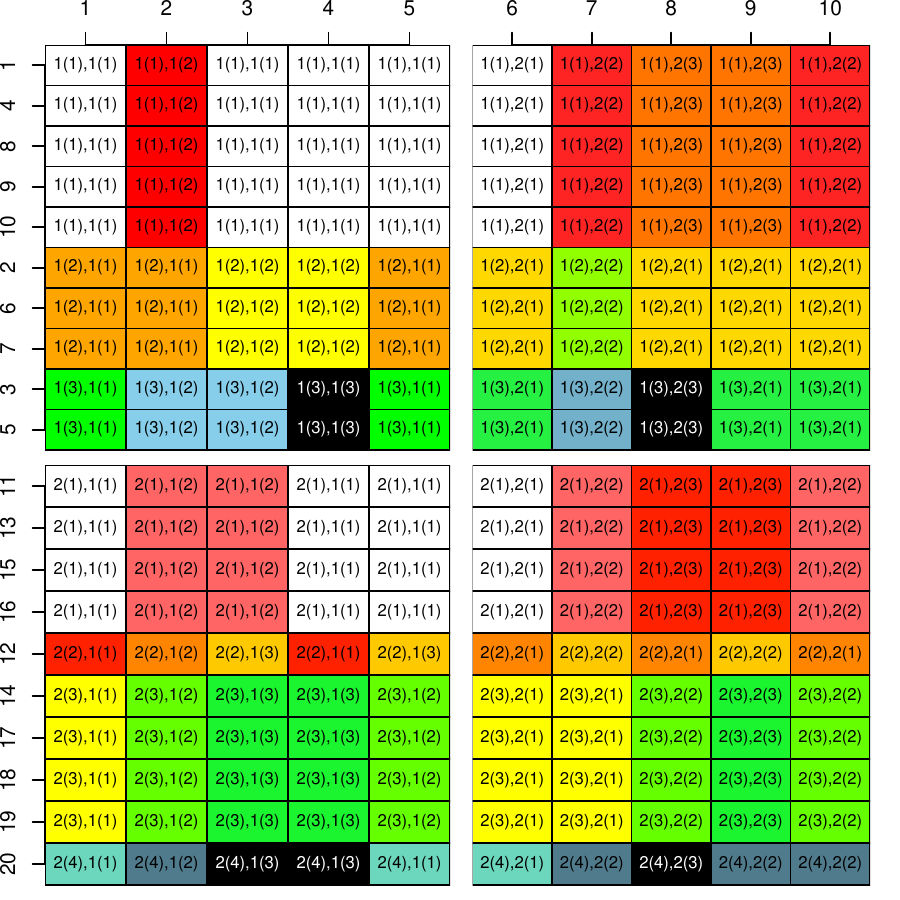}\]
    \vspace*{-0.3in}
    \[\text{(b)}\quad \includegraphics[width=2.65in]{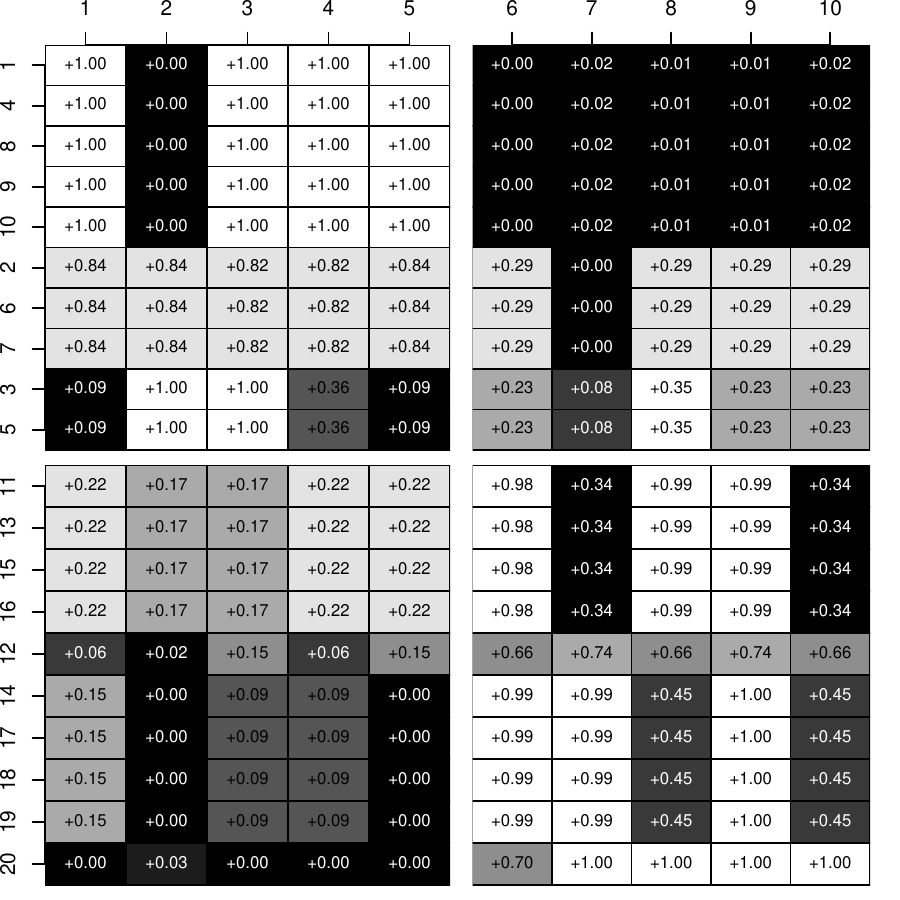}\]
    \vspace*{-0.3in}
    \[\text{(c)}\quad \includegraphics[width=2.65in]{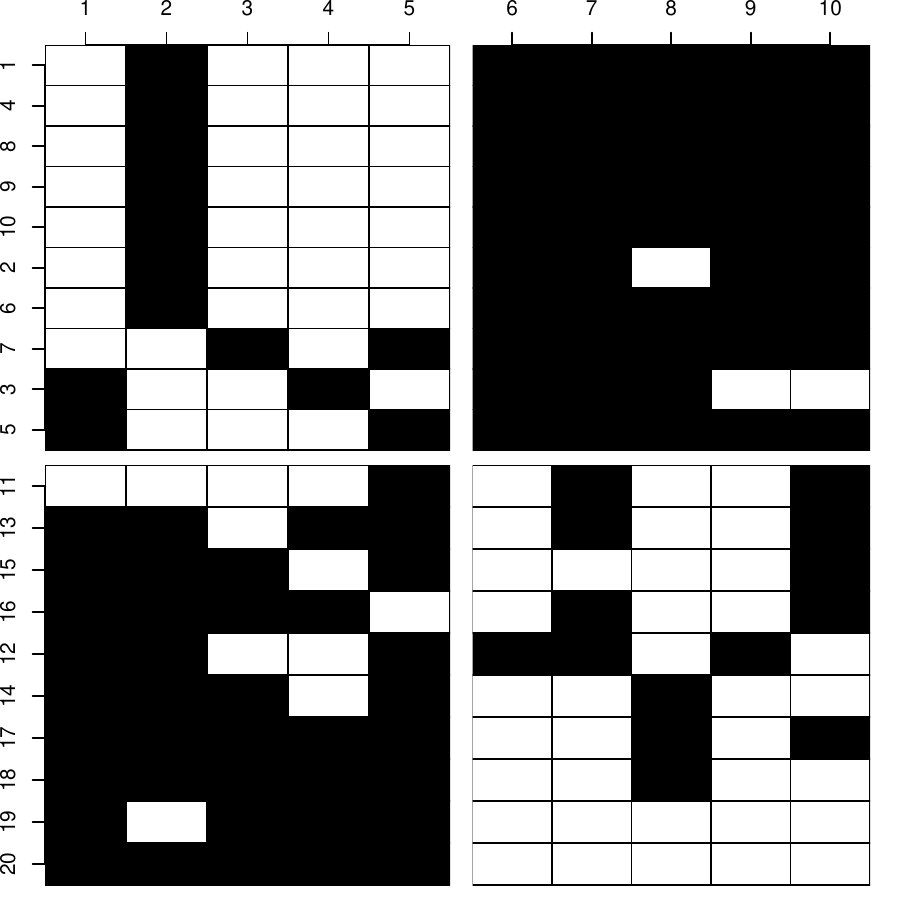}\]
    \vspace*{-0.3in}
    \caption{Simulated biclustering and simulated data: (a) Data matrix with simulated clustering labels $f(k),g(l)$ at cells $(i,j)\in C(f(k),g(l))$ (b) matrix of simulated cell success probabilities $\theta$ with the $\theta_{ij}$ value in each cell $(i,j)\in \N\times \M$; (c) simulated data, cells are white/black ($\X_{ij}=1/0$) with probability $\theta_{ij}$ given in the corresponding cell in panel (b).}
    \label{fig:simdat}
\end{figure}

Examples of $\X, \theta$ and $f(k), g(l)$-cell labels introduced in Section~\ref{sec:bicluster} are shown in Figure~\ref{fig:simdat}. 
This is an artificial example with just $F = 2$ fluid-types and $N_1 = N_2 = 10$ observations per type, simulated for the purpose of illustration. 
These are given for $F = 2$ fluid-types $\F=\{1, 2\}$, and $N=20$ profiles, $N_1=10$ with labels $\N_1=\{1,...,10\}$ for fluid-type $f=1$ and $N_2=10$ with labels $\N_2=\{11,...,20\}$ for fluid type $f=2$. 
There are $M=10$ markers with $M_1=5$ with labels $\M_1=\{1,..,5\}$ targeting the first fluid type, and $M_2=5$ with labels $\M_2=\{6,...,10\}$ targeting the second fluid type. 

Figure~\ref{fig:simdat}(a) shows a possible random realization of the biclusters. 
The row partitions in the first fluid-type are $R_1 = (R_{1(1)} = \{1, 4, 8, 9, 10\}, R_{1(2)} = \{2, 6, 7\},  R_{1(3)} = \{3, 5\})$, while in the second fluid-type $R_2 = ( R_{2(1)} = \{11, 13, 15, 16\}, R_{2(2)} = \{12\}, R_{2(3)} = \{14, 17, 18, 19\}, R_{2(4)} = \{20\})$, resulting in $K_{1} = 3$ and $K_2 = 4$. 
The column partitions in the first subtype $R_{1(1)}$ in the first fluid-type are, for $\M_1$, $S_{1(1),1} = (S_{1(1),1(1)}=\{1, 3, 4, 5\},S_{1(1),1(2)}=\{2\})$, giving $K_{1(1), 1} = 2$, whereas for $\M_2$, $S_{1(1),2} = (S_{1(1),2(1)}=\{6\},S_{1(1),2(2)}=\{7,10\},S_{1(1),2(3)}=\{8,9\})$, which means $K_{1(1),2} = 3$. 
The column partitions in the second subtype $R_{1(2)}$ in the first fluid-type are, for $\M_1$, $S_{1(2),1}=(S_{1(2),1(1)}=\{1, 2, 5\}, S_{1(2),1(2)}=\{3,4\})$, while for $\M_2$, $S_{1(2),2}=(S_{1(2),2(1)}=\{6, 8, 9, 10\},S_{1(2),2(2)}=\{7\})$, which gives $K_{1(2),1} = K_{1(2), 2} = 2$. 
We omit column partitions in the third subtype of the first fluid type and the clusters for the second fluid-type. 
As an example of bicluster notation, the bicluster $C_{2(4),2(2)} = \{(20,7), (20,9), (20, 10)\}$ (the gray-blue bicluster in the bottom right block in (a)) has $c_{2(4),2(2)} = 3$ cells.

Figure~\ref{fig:simdat} (b) shows a possible random realization of the $\theta$-values, given the biclustering. These are constant within $(f(k), g(l))$ clusters. 
For example all eight cells $(i,j)\in C_{2(1),2(2)}$ (the pink bicluster in the bottom right block of (a)) have $\theta_{i,j}=\theta^*_{2(1),2(2)}=0.34$.
Figure~\ref{fig:simdat} (c) shows a realization of the $N\times M$ binary data-matrix $\X$. 
Each entry $x_{i,j}$ is an independent Bernoulli random variable with success probability $\theta_{i,j}$. 
The number of successes $s_{f(k),g(l)}$ in bicluster $(f(k),g(l))=(2(1),1(1))$ (the white bicluster in the bottom left block in (a)) is $s_{2(1),1(1)}=4$.

\section{Choice of prior hyperparameters}\label{sec:fixing-a-and-b}
This section outlines the choice of prior hyperparameter values for analyzing our forensic datasets.
In Section~\ref{sec:theta-star-theta-prior}, the activation probabilities $\theta^*_{f(k),g(l)}$ in bicluster $(f(k),g(l))\in \C_{f,g}$ were taken to have priors $\theta^*_{f(k),g(l)}\sim \mbox{Beta}(a_{f,g},b_{f,g}),\ f,g\in F$ with potential different shape parameters in each of the 25 fluid-marker blocks $(f,g)\in \F\times \F$, which gives 50 prior hyperparameters there. 
We have a further twenty hyperparameters $\alpha_f, J_f$ and $\beta_g, L_g$ for the MCRP distributions taken in Sections~\ref{sec:row-clusters-prior} and \ref{sec:column-clusters-prior} respectively. 
We used the training data to assign these hyperparameters fixed values.

\subsection{Prior row and column clustering hyperparameters} 
The number of columns in a marker group is small ($M_g = 5$ for $g = 1,3,4,5$ and $M_2 = 7$), so we were not concerned about partitions with a long tail of small clusters and simply took $L_g = M_g$ (the number of columns in marker group $g$) with $\beta_g$ chosen to give a prior probability $\pi(L_g = 1) \approx 0.5$ for all $g\in \F$, inducing  $\beta_2=0.375$ and $\beta_g=0.49$ for $g\ne 2$. 
Following simple exploratory work on the training data ($k$-means clustering of rows and visualization in Figure~\ref{fig:dataGrid}), we set the MCRP row parameter $J_f = 5,\ f\in \F$ to give a hard upper limit on the number of fluid subtypes equal five. 
Next, we select $\alpha_f$ separately for each $f\in\F$ to give a prior probability $\pi(\alpha_f \in \{1, 2\}) \approx 0.5$, (giving $\alpha_1 = 0.6025$, $\alpha_2 = 0.725$, $\alpha_3=0.55$, $\alpha_4 = 0.585$ and $\alpha_5=0.525$) as the $N_f$-values in Table~\ref{tab:rnaProfile} vary with $f\in\F$. 
A sensitivity analysis is performed with smaller and larger $J$-values in Appendix~\ref{app:J-sensitivity-analysis}.

\subsection{Prior activation probability hyperparameters} 
Inspection of Figure~\ref{fig:dataGrid} suggests that in blocks $(f,g) \in \F\times \F$ with $f = g$, suitable choices of $a_{f,g}$'s and $b_{f,g}$ may be $a_{f,f}=1/2$ and $b_{f,f}=1/4$ and $a_{f,g}=1/2$ and $b_{f,g} = 1$ with $f \neq g$ for all $f,g \in \F$.
By inspection, we observe general patterns among candidate biclusters.
In the on-diagonal blocks (i.e., $f=g$), biclusters tend to be either completely active or inactive, while in the off-diagonal blocks (i.e., $f \neq g$), they tend to be inactive but noisier.
However, exceptions seem warranted in the on-diagonal MTB/MTB and CVF/CVF blocks and the off-diagonal MTB/CVF and MTB/BLD blocks, where $a_{1,1}=b_{1,1}=1$, $a_{2,2}=b_{2,2}=1$, $a_{2,1}=b_{2,1}=1$ and $a_{2,4}=b_{2,4}=1$, respectively.

We checked (and slightly modified) these elicited values following MCMC simulations targeting $\pi(a_f, b_f, R_f|\X_f)$ (with $a_f=(a_{f,1},\dots, a_{f, F})$ and $b=(b_{f,1},\dots,b_{f, F})$) separately for each fluid-type, taking independent $\Gamma(0.01,0.01)$ priors for all $a_{f,g}$ and $b_{f,g}$.
Posterior mean parameter values are given in Table~\ref{tab:a-and-b-post}. 
\begin{table}[htb]
  \centering
  \begin{tabular}{r|ccccc}
       &\multicolumn{5}{c}{Marker group}\\
    Fluid-type & CVF & MTB & SLV & BLD & SMN 
    \\[0.05in] \hline \\[-0.1in]
    CVF & $(1.1,0.5)$ & $(0.5,2.5)$ & $(0.1,0.9)$ & $(0.1,1.6)$ & $(0.3,1.0)$ \\
    MTB & $(1.8,0.6)$ & $(0.6,1.0)$ & $(0.8,19)$ & $(0.8,0.9)$ & $(0.4,13)$ \\
    SLV & $(0.1,0.5)$ & $(0.3,7)$ & $(0.3,0.1)$ & $(1e\mhyphen 4,4.5)$ & $(0.8,40)$ \\
    BLD & $(0.4,43)$ & $(0.1,0.8)$ & $(0.1,10)$ & $(0.4,0.1)$ & $(1e\mhyphen 3,4.9)$ \\
    SMN & $(0.2,15)$ & $(0.4,8.1)$ & $(0.2,3.5)$ & $(0.2,35)$ & $(0.6,0.2)$
  \end{tabular}
  \caption{The grid gives posterior mean values of prior parameters $(a_{f,g},b_{f,g})$ for the $\mbox{Beta}(a_{f,g},b_{f,g})$-prior distributions for activation probabilities $\theta^*_{f(k),g(l)}$ in bicusters $(f(k),g(l))\in \C_{f,g}$ across the twenty five fluid-type/marker-group blocks $(f,g)\in \F\times\F$. 
  }
  \label{tab:a-and-b-post}
\end{table}
We could of course use this in our overall analysis, but we found mixing rather slow (even on the smaller separate fluid-type data sets) and preferred to fix these hyperparameters.
This choice was borne out by the well-calibrated posterior distributions we subsequently obtained for the test data in Section~\ref{sec:results}. 
All the training data in Table~\ref{tab:rnaProfile} were used for these $a$-and-$b$ measurements and no test data and no unlabeled profiles, so $\U=\emptyset$ and $\N_f=\T_f$ here.

\begin{figure}[ht]
    \includegraphics[width=10.5cm]{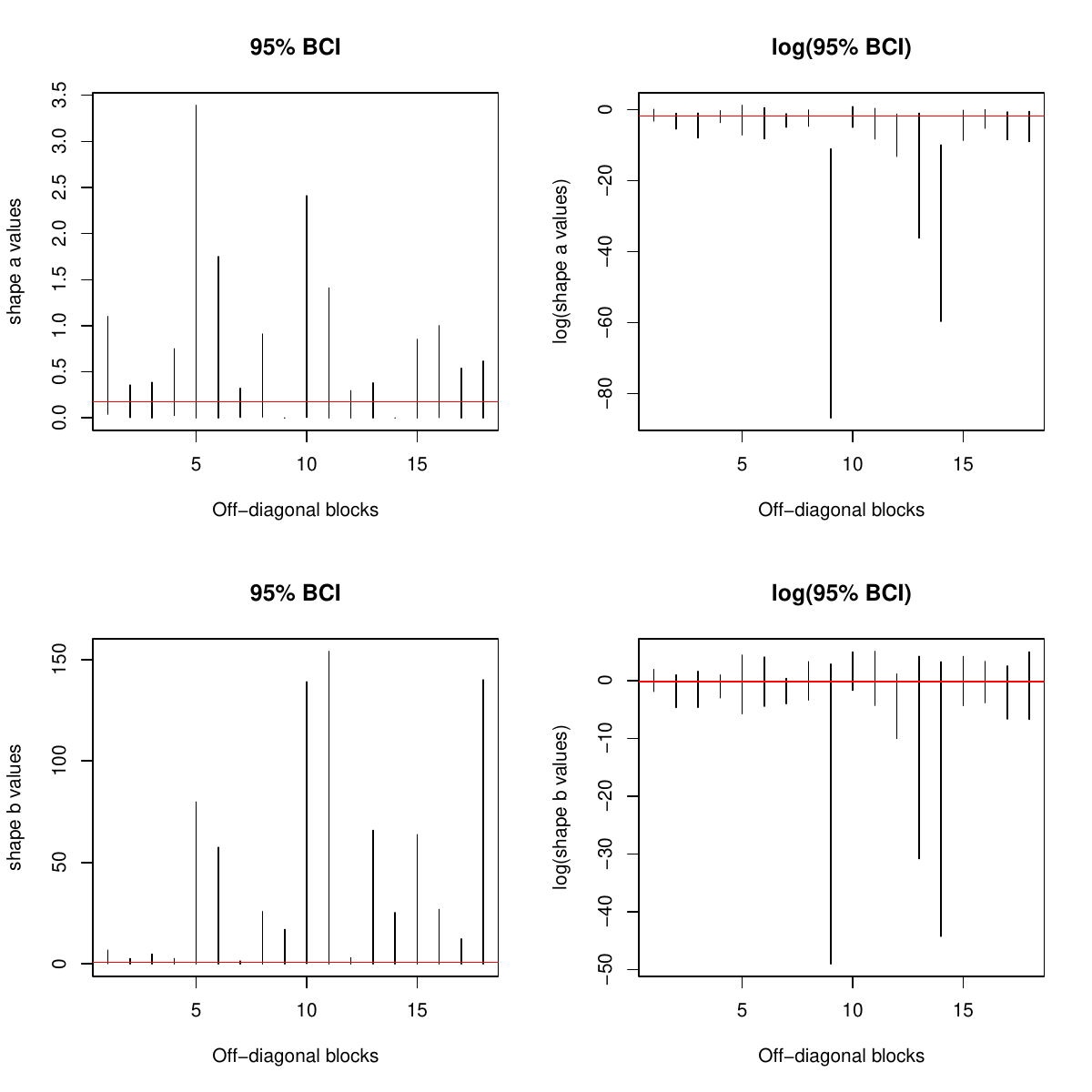}
    \centering
    \caption{The graphs show 95\% HPD Bayesian credible intervals for the shape parameters in each of the eighteen off-diagonal blocks excluding MTB/CVF and MTB/BLD. Left column shows $a$ (top) and $b$ (bottom) intervals, right column shows $\log(a)$ (top) and $\log(b)$ (bottom).}
    \label{fig:2022_09_21_ff_diag_ab_95_BCI}
\end{figure}

Marginal Highest Posterior Density (HPD) Bayesian Credible Intervals (BCI) for each of the eighteen off-diagonal blocks (excluding MTB/CVF and MTB/BLD) are given (on the original and log scales) in Figure~\ref{fig:2022_09_21_ff_diag_ab_95_BCI}. 
The horizontal red lines gives $a$-and-$b$ values at $a_{f,g} = 0.174$ and $b_{f,g} = 0.807$. 
These are consensus values for all distributions except $a_{3,4}$ (SLV/BLD, $9$'th BCI) and  $a_{4,5}$ (BLD/SMN, $14$'th BCI).

Since we expect more noise in the test data (and in real casework data) than in the training data, we use slightly ``less extreme'' values than the training data suggest, especially for blocks such as BLD/SMN ($f = 4, g = 5$) and SLV/BLD ($f = 3, g = 4$), which with all marker signal absent in the training data. 
\begin{table}[htb]
  \centering
  \begin{tabular}{r|ccccc}
       &\multicolumn{5}{c}{Marker group}\\
    Fluid-type & CVF & MTB & SLV & BLD & SMN 
    \\[0.05in] \hline \\[-0.1in]
    CVF & $(1,1)$ & $(0.2,0.8)$ & $(0.2,0.8)$ & $(0.2,0.8)$ & $(0.2,0.8)$ \\
    MTB & $(1,1)$ & $(1,1)$ & $(0.2,0.8)$ & $(1,1)$ & $(0.2,0.8)$ \\
    SLV & $(0.2,0.8)$ & $(0.2,0.8)$ & $(0.45,0.15)$ & $(0.2,0.8)$ & $(0.2,0.8)$ \\
    BLD & $(0.2,0.8)$ & $(0.2,0.8)$ & $(0.2,0.8)$ & $(0.45,0.15)$ & $(0.2,0.8)$ \\
    SMN & $(0.2,0.8)$ & $(0.2,0.8)$ & $(0.2,0.8)$ & $(0.2,0.8)$ & $(0.45,0.15)$
  \end{tabular}
  \caption{The grid gives the elicited fixed values of the prior hyperparameters $(a_{f,g},b_{f,g})$ for the $\mbox{Beta}(a_{f,g},b_{f,g})$-prior distributions for activation probabilities $\theta^*_{f(k),g(l)}$ in biclusters $(f(k),g(l))\in \C_{f,g}$ across the 25 fluid-type/marker-group blocks $(f,g)\in \F\times\F$.}
  \label{tab:a-and-b-chosen}
\end{table}
The chosen fixed values are given in Table~\ref{tab:a-and-b-chosen}, which are generally more conservative than the values in Table~\ref{tab:a-and-b-post} would allow, but close to the consensus values from Figure~\ref{fig:2022_09_21_ff_diag_ab_95_BCI} for off-diagonal blocks.

\section{MCMC updates}\label{sec:mcmc-all}
We describe the various proposal distributions used in the MCMC to sample the posterior distribution under the nested finite mixture model. 
These target $\pi(R|\X,y_\T) = \prod_{f\in F}\pi(R_f|\X_f,y_{\T_f})$ for $R\in\Xi^{\, *}_{\U|\T}$, the posterior including unlabeled profiles in Section~\ref{sec:post_missing_fluid_types}. 
Since the knowledge of $R$ determines $\V$ exactly, we can equivalently write the target $\pi(R,\V|\X,y_\T)$ for $\V\in \Psi_\U$ and $R\in \Xi^{*}_{\T,\V}$.
For each proposal move, we first outline the general algorithm, followed by the mathematical details.

\subsection{Sampling the subtype assignment with fixed fluid-type assignments} \label{sec:MH_fixed_FT}

\subsubsection{Outline}
A value $i \in \N$ is chosen uniformly at random.
The set of potential subtypes consists of the existing subtypes, except the current subtype containing $i$.
If the current number of subtypes is less than $J$ and $i$ does not form a singleton subtype in its current fluid-type, an empty subtype is added to the potential set of subtypes.
A proposed subtype is randomly selected from the potential set of subtypes, followed by accepting or rejecting this proposal based on the appropriate Metropolis-Hastings ratio.

\subsubsection{Details} 
Assuming the current state of the chain is $R\in\Xi^{\, *}_{\T,\V}$ with fixed assignment partition $\V$, the subtype partition $R_f$ for all $f \in \F$ is updated according to the following. 
We give the general setting (where for example, a fluid-type could have just one profile), so in practice, some of the ``corner cases'' we accommodate explicitly below will not arise on our data.

Firstly, we select at random $f \sim U(\F\setminus\F^{(1)})$, where 
$\F^{(1)} = \left\{ \tilde{f} \in \F: N_{\tilde{f}} = 1 \right\}$.
This setup forbids updating subtypes within a fluid-type containing only a single mRNA profile, as it can only have one subtype.
Then, we randomly draw an existing subtype $k \sim U\left\{1, ..., K_f\right\}$, from which a profile $i \sim U(R_{f(k)})$.

Let $n_{f(k)} = |R_{f(k)}|$ give the number of mRNA profiles in $R_{f(k)}$. 
Suppose (Case A) that $R_{f(k)}$ is a singleton subtype, i.e., $R_{f(k)} = \{i\}$, and hence $n_{f(k)} = 1$. 
Select a new subtype $k'\sim U(\{1,...,K_f\}\setminus\{k\})$ of fluid-type $f$, so that $k'\ne k$. 
Subtype $f(k)$ will be removed if the move to the new subtype is accepted. 
Note that $K_f$ must be at least two in Case A as we have a singleton and $N_f>1$ so there is another profile and hence a subtype with $k'\ne k$.
In other proposals (like Case B below), $k'$ is chosen from a set including a new $K_f+1$ subtype. 
That choice is not included in Case A, as it would be a pointless singleton-to-singleton move. 
Re-indexing the subtypes in fluid type $f$, we set
$$
R'_{f(\ell)}=\begin{cases}
			R_{f(\ell)}, & \text{for } \ell \in \{1,...,k-1\}\\
            R_{f(\ell+1)}, & \text{for } \ell \in \{k,...,K_f-1\}
		 \end{cases}
$$

Finally, we set $R'_{f(k')}\leftarrow R'_{f(k')}\cup\{r\}$, and hence the proposed subtype partition of fluid-type $f$ is $R'_{f} = (R'_{f(1)},\dots,R'_{f(K_f-1)})$.
The proposal probability for this update is
\[
q_A(R'_f|R_f)=\frac{1}{K_f (K_{f}-1)}
\]

If (Case B) $n_{f(k)} > 1$, then $R_{f(k)}$ is not a singleton subtype, so it remains non-empty after the removal of profile $i$. 
We randomly select a new subtype 
$$
k' \sim \begin{cases}
			U(\{1,...,K_f+1\}\setminus\{k\}), & \text{if } K_f < J_f,\\
            U(\{1,...,K_f\}\setminus\{k\}), & \text{if } K_f = J_f.
		 \end{cases}
$$
Again, the selection procedure ensures $k'\ne k$, and if $k' = K_f + 1$, then a new (singleton) will be created. 

Suppose (Case B.1) $k'\le K_f$, which means that the proposed subtype already exists, and the total number of subtypes of fluid-type $k$ will not be changed by the update.
The proposed subtype partition for fluid-type $f$ is $R'_{f} = (R'_{f(1)},\dots,R'_{f(K_f-1)})$, with
$$
R'_{f(\ell)}=\begin{cases}
	R_{f(\ell)}\setminus\{r\} & \text{for } \ell = k,\\
    R_{f(\ell)}\cup\{r\} & \text{for } \ell = k',\\
    R_{f(\ell)} & \text{otherwise}.
\end{cases}
$$
On the other hand, if (Case B.2) $k' = K_f + 1$, resulting in a new subtype for the proposal, then proposed subtype partition for fluid-type $f$ is $R'_f = R_f \cup \{R_{f(K_f+1)}\}$ with $R_{f(K_f+1)}=\{i\}$. 
These two updates induce the proposal probability 
$$
q_{B}(R'_f|R_f) =\begin{cases}
    \frac{1}{K^2_f n_{f(k)}}& \text{for } K_f < J_f,\\
    \frac{1}{K_f(K_f - 1)\, n_{f(k)}} & \text{for } K_f = J_f.
\end{cases}
$$
Note that for Case B.2, we must have $K_f < J_f$.

Detailed balance holds between Case A and Case B.2. 
The acceptance probability for the proposed state $R'_f$ in Case A is
\begin{align*}
    \alpha_A(R'_f|R_f)&=\left\{1,\ \frac{\pi(R'_f|\X_f)\,q_B(R_f|R'_f)}{\pi(R_f|\X_f)\,q_A(R'_f|R_f)}\right\}\\
    &=\left\{1,\ \frac{\pi_R(R'_f)\prod_{\ell=1}^{K_f-1} p(\X_{f(\ell)}|R'_{f(\ell)})\, K_f (K_{f}-1)}{\pi_R(R_f)\prod_{\ell=1}^{K_f} p(\X_{f(\ell)}|R_{f(\ell)})(K_f-1)^2 (n_{f(k')} + 1)}\right\}\\
    &=\left\{1,\ \frac{\pi_R(R'_f)\, p(\X_{f(k')}|R'_{f(k')})\, K_f }{\pi_R(R_f)\, p(\X_{f(k')}|R_{f(k')})\, p(\X_{f(k)}|R_{f(k)})(K_f-1) (n_{f(k')} + 1) }\right\}.
\end{align*}
On the other hand, the acceptance probability for the proposed state $R'_f$ for Case B.1 is
\begin{align*}
  \alpha_{B.1}(R'_f|R_f) &=\left\{1,\ \frac{\pi_R(R'_f)\,
  p(\X_{f(k')}|R'_{f(k')})\, p(\X_{f(k)}|R'_{f(k)})
  n_{f(k)}   }
    {\pi_R(R_f)\, p(\X_{f(k')}|R_{f(k')})\, p(\X_{f(k)}|R_{f(k)}) (n_{f(k')} + 1) }\right\},
\end{align*}
while that for Case B.2 is
\begin{align*}
    \alpha_{B.2}(R'_f|R_f)&=\left\{1,\ \frac{\pi_R(R'_f)\, p(\X_{f(k')}|R'_{f(k')})\, p(\X_{f(k)}|R'_{f(k)})\, n_{f(k)}\, K_{f}  }
    {\pi_R(R_f)\, p(\X_{f(k)}|R_{f(k}) (K_f + 1)  }\right\}.\\
\end{align*}

Since $\V$ is fixed, and independent MCRP is applied to the subtype clustering to each fluid-type $f \in \F$, these updates can be applied in parallel to target each $\pi(R_f|\X_f,\V_f),\ f\in \F$ independently and sample $\pi(R|\X, \V),\ R\in \Xi^*_{\T,\V}$. 

The following section provides an update on mixing over $\V$.

\subsection{Metropolis-Hastings sampler for updates across fluid-types} \label{sec:mcmc-between-types}

\subsubsection{Outline}

The index of an unlabeled type $i \in \U$ is selected uniformly at random.
Uniformly choose a subtype $f' \in \F_{-i}$ at random, where $\F_{-i} = \F \setminus y_i, y_i \in \F$.
We then randomly and uniformly select a subtype from the chosen type above.
The potential subtypes include the existing subtypes in fluid-type $f'$, but also include an empty subtype if $|f'| < J$.
The proposed fluid-type and subtype for $i$ are accepted or rejected based on the appropriate Metropolis-Hastings ratio.

\subsubsection{Details}
Let $R\in\Xi^{\, *}_{\U|\T}$ be the current state of the Markov Chain. 
As remarked above, if $R$ is given, then the corresponding $\V$ 
is determined. 
We choose an unlabeled profile $i \sim U(\U)$ at random, i.e.,  $i \in \{N + 1, \dots, N + U\}$.
Suppose $i \in R_{f(k)}$ for some $k\in \{1,...,K_f\}$. 
Subsequently, we randomly select $f'\sim U(\F\setminus\{f\})$, draw
$$
k' \sim\begin{cases}
    U\{1,...,K_{f'}+1\} & \text{for } K_{f'} < J_{f'},\\
    U\{1,...,K_{f'}\}  & \text{for } K_{f'} = J_{f'}
\end{cases}
$$
and remove $i$ from $R_{f(k)}$ and add it in $R'_{f'(k')}$.
If $k'\le K_{f'}$ then $i$ is added to an existing subtype $R_{f(k')}$ of fluid-type $f'$, whereas if  $k' = K_{f'} + 1$ then $i$ forms a new subtype $R_{f(K_{f'}+1)} = \{i\}$. 



If (Case A) $R_{f(k)} = \{i\}$, the update removes subtype $R_{f(k)}$ from fluid-type $f$, so the proposed partition for fluid-type $f$ is $R'_f = \cup_{\tilde{k} = 1}^{K'_f}\{R'_{f(\tilde{k}) }\}$, where

$$
R'_{f(\tilde{k}) }=\begin{cases}
    R_{f(\tilde{k}) } & \text{for } \tilde{k} \in \{1,..., k - 1\},\\
    R_{f(\tilde{k} + 1)} & \text{for } \tilde{k} \in \{k,..., K'_f = K_f - 1\}.
\end{cases}
$$

If (Case B) $n_{f(k)}>1$, then $R_{f(k)}$ is not a singleton, so $K_f$ remain unchanged and
$$
R'_{f(\tilde{k}) }=\begin{cases}
    R_{f(\tilde{k}) } & \text{for } \tilde{k} \in \{1,..., K_f\}\setminus k,\\
    R_{f(\tilde{k})}\setminus\{i\} & \text{for } \tilde{k} = k.
\end{cases}
$$

For both Cases A and B,
if  $k' = K_{f'} + 1$, implying $K_{f'} < J_{f'}$, then a new subtype is created in the proposed fluid-type $f'$. 
We obtain $R'_{f'} = \cup_{\tilde{k} = 1}^{K'_{f'}}\{R'_{f'(\tilde{k}) }\}$ with
$$
R'_{f'(\tilde{k}) }=\begin{cases}
    R_{f'(\tilde{k}) } & \text{for } \tilde{k} \in \{1,..., K_{f'}\},\\
    \{i\} & \text{for }  \tilde{k}=K'_{f'} = K_{f'} + 1.
\end{cases}
$$
On the other hand, if $k' \le K_{f'}$, that is, the proposed subtype assignment already exists, then we have
$$
R'_{f'(\tilde{k}) }=\begin{cases}
    R_{f'(\tilde{k}) } & \text{for } \tilde{k} \in \{1,..., K_{f'}\}\setminus k',\\
    R_{f'(k')} \cup \{i\} & \text{for } \tilde{k} = k',
\end{cases}
$$
which is permitted for $K_{f'} \leq J_{f'}$.

Depending on $K_{f'}$ only, the proposal probabilities are 
\[q(R'|R)=\begin{cases}
(U (F-1)\, (K_{f'}+1))^{-1} & \text{if } K_{f'}< J_{f'},\\
(U (F-1)\, K_{f'})^{-1} & \text{if } K_{f'}= J_{f'}.
\end{cases}\]

In Case A with $R_{f(k)} = \{i\}$, the Hasting's ratio $q(R|R')/q(R'|R)$ is
$$
H_A(R'|R) =\begin{cases}
    \frac{K_{f'} + 1}{K_f} & \text{if } K_{f'} < J_{f'},\\
    \frac{K_{f'}}{K_f} & \text{if } K_{f'} = J_{f'}.
\end{cases}
$$
When $n_{f(k)} > 1$, for Case B.1 with $K_{f} < J_{f}$, the Hasting's ratio is
$$
H_{B.1}(R'|R) =\begin{cases}
    \frac{K_{f'} + 1}{K_f + 1} & \text{if } K_{f'} < J_{f'},\\
    \frac{K_{f'}}{K_f + 1} & \text{if } K_{f'} = J_{f'},
\end{cases}
$$
whereas for Case B.2 with $K_{f} = J_{f}$, the Hastings ratio is simply $H_{B.2}(R'|R)=H_A(R'|R)$.

\subsection{Gibbs sampler for updates across fluid-types in the Cut-Model inference}\label{sec:mcmc-gibbs}

\subsubsection{Outline}
The Gibbs Sampler presented here is a special case of a Nested MCMC sampler. Nested MCMC for Cut-Models was suggested in \cite{Plummer2015CutsModels} and is described above in Section~\ref{sec:cut-model}, where some of the following notation is introduced. 
Our sampler proposes a new partition by randomly selecting the fluid-type and subtype for an unlabeled profile $i$ according to the full conditional probability $p(R'|R_{-i})$. 
Details below explain the sampling algorithm for classifying a single profile (SPC) and multiple profiles simultaneously (JPC).

\subsubsection{Details}
After obtaining $\pi(Q|\X_\T)$ using MCMC and the sampler in Section ~\ref{sec:MH_fixed_FT}, for each $t = 1,...,T_0$, Gibbs sampling is employed to simulate a side chain from $\pi(R|\X,Q^{(t)})$.
For the $t$th side chain, we first randomly select $R^{(t)}_0$, an initial state for the row partitions of $(\T,\U)$, conditioned on $Q^{(t)}$ such that $R^{(t)}_0 \in \Xi^{\, *}_{\U|\T}(Q^{(t)})$. 
Therefore, the initial joint partition $R^{(t)}_0$ of the labeled and unlabeled data contains the partition $Q^{(t)}$ of the labeled data as a joint sub-partition.

At step $t'=1,\dots, T_1$ of this MCMC chain, we randomly select an unlabeled profile $i \sim U(\U)$.
Subsequently, we construct a Gibbs sampler to draw $R^{(t, t')} \in \Xi^{\, *}_{\U|\T}(Q^{(t)}) $ from the full conditional distribution 
$$
R^{(t, t')} \sim \pi\left(\cdot|R^{(t, t' - 1)}_{-i}\right),
$$
which is equivalent to sampling the vector of probabilities with components given by
$$
    \Pr\left(i \in R^{(t, t')}_{f(k)}|R^{(t,t' - 1)}_{-i}\right) \propto p\left(\X|R^{(t, t'-1) }_{-i}, i \in R^{(t,t')}_{f(k)}\right)\pi\left(R^{(t,t')}\right),
$$
for $f \in \F$ and 
$$
k \in \begin{cases}
     \{1, ..., K^{(t,t')}_f + 1\} & \text{if } K^{(t,t')} < J_{f},\\
    \{1, ..., K^{(t,t')}_f \} & \text{if } K^{(t,t')} = J_{f},
\end{cases}
$$
where $K^{(t,t')}_f$ is the number of subtypes in fluid-type $f \in \F$ in $R^{(t,t')}$.
When the analyses are performed with SPC, $T_1 = 1$ is sufficient for all $t \in T_0$; in other words, a single draw by the Gibbs sampler is sufficient for each subchain.
However, in the JPC analyses, for each proposal by the Gibbs sampler, we select an unlabeled profile $i \sim \text{Uniform}(\U)$ at random and draw $R'$ from $p(R'|R_{-i})$. In this case, length of each subchain $T_1$ is set such that each unlabeled profile $i \in \U$ would be updated $\ge 10$ times on average.


\subsection{Runtime, Convergence and Implementation Checks for Metropolis Hastings}\label{sec:mcmc-convergence}
To get an idea of runtimes, in the best case, the Cut-Model setup estimating a fluid-type for a single profile, we get around 10 Effective Independent Samples per second (so 600 EIS'/minute) on a standard office desktop computer in a careful Java implementation, but without parallelization. 
This value is computed using the Effective Sample Size (ESS) for the fluid-type of the unlabeled profile and excludes the one-off cost of sampling subtypes for training data. 
In the worst case, Bayesian inference for joint estimation of fluid-type for our test set of 46 samples, which includes estimating the subtype structure for the 321 profiles in the training data, we obtain approximately 1.5 EIS'/minute, equivalently, roughly a half-day to get the joint fluid-type posterior for 46 profiles if we target an ESS equal 1000 for each profile. 
This is just representative, using a central ESS value among those we observed for sample fluid-type values $y_i=\{f: i\in \V_f\}$. 

For convergence diagnostics, we focus on achieving large ESS values for the random fluid-type $y_i$ of unlabeled samples $i\in \U$. 
Across all 740 MCMC runs for the different analyses in this paper (the majority being LOOCV checks on training data) an ESS well above 1000 seems typical. 
The ESS values and ranges for all the analyses are given in Table~\ref{tab:analysisSummaryESS}.
\begin{table}
 \caption{Range of the ESS of the mode type variable across all unlabelled profiles for each analysis. The mode type variable is coded in binary format, where $1 = \text{yes}$ and $0 = \text{no}.$
 The ESS range is calculated excluding any unlabelled profiles with a mode type that has a posterior probability of 1.0. 
 Below SPC and JPC stand for ``Single'' and ``Joint'' Profile Classification, respectively.}
\begin{tabular}{ |c|| c| l | c | }
 \hline
\multirow{2}{*}{Analysis}  & ESS Median & \multicolumn{1}{c|}{\multirow{2}{*}{Purpose}} & \multirow{2}{*}{Section}\\
 & \& Range  & & \\
 \hline
\multirow{3}{*}{\shortstack[c]{LOOCV on\\the training set in\\Bayesian inference}} & \multirow{3}{*}{\shortstack[c]{4500\\(2772, 4500)} } & Evaluate goodness-of-fit, calibration and &  \multirow{8}{*}{Section ~\ref{sec:loocv}} \\
 &     & classification performance with Bayesian &\\
&     & inference, using model and lab-quality data.&\\
\cline{1-3}
\multirow{5}{*}{\shortstack[c]{LOOCV on\\the training set in\\Cut-Model inference}}    & \multirow{5}{*}{\shortstack[c]{4500\\(2412, 4500)}} & Evaluate goodness-of-fit, calibration and &\\
 &   &  classification performance with Cut-Model &\\
 &   &  inference, using model and lab-quality data, &\\
 &   &  and compare to the corresponding Bayesian. &\\
  &   &  analysis. &\\
\hline 
\multirow{3}{*}{\shortstack[c]{JPC of profiles\\in the test set in\\Bayesian inference}}  & \multirow{3}{*}{\shortstack[c]{3955\\(1672, 4500)} } & Evaluate the performance of the trained model &  \multirow{18}{*}{Section ~\ref{sec:classTestSet}}\\
&   &  for jointly classifying casework-like data using &\\
 &   &  Bayesian inference and lab-quality training data. &\\
\cline{1-3}
\multirow{5}{*}{\shortstack[c]{JPC of profiles\\in the test set in\\Cut-Model inference}}  &   \multirow{5}{*}{\shortstack[c]{4500\\(2402, 4500)} } & Evaluate the performance of the trained model &\\
 &   &  for jointly classifying casework-like data using &\\
 &   &  Cut-Model inference and lab-quality training  &\\
 &   &  data, and compare to the corresponding &\\
  &   &  Bayesian analysis. &\\
\cline{1-3}
\multirow{5}{*}{\shortstack[c]{SPC of profiles\\in the test set in\\Bayesian inference}}  & \multirow{5}{*}{\shortstack[c]{4500\\(1313, 4500)} } & Evaluate the performance of the trained model &  \\
&   &   for classifying casework-like data one at a time &\\
 &   &  using Bayesian inference and lab-quality training &\\
 &  &  data, and compare to the corresponding joint &\\
  &  &  analysis. &\\
\cline{1-3}
\multirow{5}{*}{\shortstack[c]{SPC of profiles\\in the test set in\\Cut-Model inference}} & \multirow{5}{*}{\shortstack[c]{4500\\(936, 4500) }} & Evaluate the performance of the trained model &\\
 &   &  for classifying casework-like data one at a time &\\
 &   &   using Cut-Model inference, and lab-quality &\\
 &   &  training data, and compare to the corresponding &\\
 &   &  Bayesian analysis. &\\
\hline 
\multirow{4}{*}{\shortstack[c]{JPC of profiles\\in the test set in\\Bayesian inference\\($J_f = 1$ \& $L_g = 1$) }}   & \multirow{4}{*}{\shortstack[c]{4500\\(3117, 4500)} } &  \multirow{16}{*}{\shortstack[c]{To investigate the effect of cluster bounds on\\the classification of casework-like data.} } &  \multirow{16}{*}{Appendix ~\ref{app:J-sensitivity-analysis}}\\
&   &  &  \\
&   &  & \\
&   &  & \\
\cline{1-2}
\multirow{4}{*}{\shortstack[c]{JPC of profiles\\in the test set in\\Bayesian inference\\($J_f = 1$ \& $L_g = M_g$) }}     & \multirow{4}{*}{\shortstack[c]{4500\\(3496, 4500)} } & &\\
&   &  &  \\
&   &  & \\
&   &  & \\
\cline{1-2}
\multirow{4}{*}{\shortstack[c]{JPC of profiles\\in the test set in\\Bayesian inference\\($J_f = 10$ \& $L_g = M_g$) }}    & \multirow{4}{*}{\shortstack[c]{3703\\(1856, 4500)} } & &\\
&   &  & \\
&   &  & \\
 &   &  & \\
\cline{1-2}
\multirow{4}{*}{\shortstack[c]{JPC of profiles\\in the test set in\\Bayesian inference\\($J_f = 15$ \& $L_g = M_g$) }}   & \multirow{4}{*}{\shortstack[c]{3765\\(1277, 4500)} } & &\\
&   &  & \\
&   &  & \\
 &   &  & \\
\hline 
\end{tabular}
\label{tab:analysisSummaryESS}
\end{table}

The implementation is checked by verifying our software reproduced the MCRP distributions when setting the log-likelihood function to return zero. 
We hand-checked likelihood calculations for a small number of profiles in a single subtype. 
Also, we performed classification on the unlabeled profile with all missing values and checked that the posterior distribution for its fluid-type was indeed uniform on $\F$. 
This is discussed further in Appendix~\ref{sec:missing-data-appendix}. 
Our LOOCV checks on the training data check correct (but blinded) classification of training data with known labels. 

\section{Missing data and unbiased fluid-type labeling}
\label{sec:missing-data-appendix}

There are no missing marker values $x_{i,j},\ i\in \N,\ j\in \M$ in the training and test data.
However, if a profile did have missing marker values, then this is easy to handle as explained in Section~\ref{sec:missing-data-main}: in the product over ${(i,j)\in C(f(k),g(l))}$ in Equation~\ref{eq:base-lkd-cs}, we simply omit from the product any cells $(i,j)$ for which the value of $x_{i,j}$ is missing.

The prior $\pi_\V(\V) = F^{-U}$ on the fluid-type assignment partition $\V$ of unlabeled profiles, presented in Section~\ref{sec:post_missing_fluid_types}, takes the prior for the fluid-type of each unlabeled profile to be independent of the fluid-types of other unlabeled profiles and uniform on $\F$. 
However, it may be of concern that the labeled training data somehow distort this and changes the prior weighting, perhaps because the number of training profiles varies from one fluid-type to another, and the number of subtypes is upper-bounded. 
We formalize this by considering the posterior distribution of $\V$ when {\it all} the data $x_i,\ i\in \U$ are missing: when $\X_{\U}$ provides no information about the fluid-types of unlabeled profiles we require $\pi(\V|X_\T,\X_\U)=F^{-U}$.

A joint partition $R\in \Xi^{\, *}_{\U|\T}$ of all profiles specifies the assignment $\V$ of unlabeled profiles to fluid-types. 
However, it contains more information as each unlabeled profile $i\in \V_f$ is assigned some subtype $k\in \{1,\dots,K_f\}$, so $i\in R_{f(k)}$. 
Let $Q(R)$ be the joint partition on training profiles, defined in Section~\ref{sec:cut-model}, which we get by removing all the floating profiles from $R$. 

Consider the likelihood $p(\X|R)$ in Equation~\ref{eq:missing_type_post} with $\X=(\X_\T,\X_U)$. In this case, with all $\X_\U$ missing, the entries in $\X_\U$ do not contribute to $c$ and $s$ in Equation~\ref{eq:base-lkd-cs}. 
Consequently, the likelihood is the probability for $\X_\T$ only, and this depends only on the joint partition $Q$ of the labeled profiles. 
It follows that
\[
p(\X|R)=p(\X_{\T}|Q),
\]
where $Q=Q(R)$ as in a Cut-Model. 

The outcome $\{Q,\V\}$ occurs iff $R\in \X^{\, *}_{\T,\V}(Q)$ occurs, where $\X^{\, *}_{\T,\V}(Q)$, defined in Equation~\ref{eq:cut-R-space-given-QV}, is the set of joint partitions $R$ that contain $Q$ and assign unlabeled profiles to fluid-types according to $\V$. 
We have
\begin{align*}
    \pi(Q,\V|\X,y_\T)
    &\propto\sum_{R\in \X^{\, *}_{\T,\V}(Q)} p(\X|R)\pi_R(R)\\
    &=p(\X_{\T}|Q) \sum_{R\in \X^{\, *}_{\T,\V}(Q)} \pi_{R}(R)\\
    &=p(\X_{\T}|Q)\pi_{Q,\V}(Q,\V),
    \end{align*}
with $\pi_{Q,\V}(Q,\V)=\pi_{Q}(Q)\pi_{\V}(\V)$, so
$\pi(\V|\X,y_\T)=\pi_\V(\V)=F^{-U}$. This result is intuitive: no data, no classification. 
The same property, $\pi_{cut}(\V|\X,y_\T)=\pi_\V(\V)$, holds in the Cut model case.
We used this property to check our code.

The acceptance probabilities for move type $T\in \{A, B.1, B.2\}$ are
\begin{align*}
\alpha_T(R'|R)&=\min\left\{1,\ \frac{\pi(R'|X)}{\pi(R|X)}H_{T}(R'|R)\right\}.
\end{align*}
There is quite a bit of cancellation in the acceptance probability.
Our implementation always works with the full sum of log-likelihoods across a fluid-type, but in this sum, only the log-likelihoods of subtypes that change in the update are themselves updated. 
This is the time-limiting step and is handled efficiently.

\section{NoB-LoC-CaRMa}

Working with the transpose of NoB-LoC \citep{lee13}, we column-extend NoB-LoC \citep{lee13} to share a common row clustering 
across multiple matrices with independent column clustering in each matrix $\times$ row-partition. 
This handles biclustering of a single fluid type matrix $\X_f$. 
We then further extend this to handle multiple independent column-extended NoB-LoC processes with random row content. 
This handles biclustering $\X = (\X_f)_{f\in\F}$ as well as class assignment for unlabeled profiles. 
Recall that NoB-LoC and the BDP applied to a single matrix have the same biclustering partition process, but assign parameters $\theta$ within a bicluster in different ways (see Section \ref{sec:relDP}).

In summary, the marginal BDP likelihood for the data in a bicluster (integrated over $\theta^*$) is 
\begin{align*}
   p(\X_{f(k),g(l)}|R_{f(k)},S_{f(k),g(l)})&=  \frac{\B(a_{f,g}+s_{f(k),g(l)},b_{f,g}+c_{f(k),g(l)}-s_{f(k),g(l)})}{\B(a_{f,g},b_{f,g})}.
\end{align*}
This is replaced in NoB-LoC-CaRMa by 
\begin{align*}
   p(\X_{f(k),g(l)}|R_{f(k)},S_{f(k),g(l)})&=\prod_{i\in R_{f(k)}} \frac{\B(a_{f,g}+s_{f(k,i),g(l)},b_{f,g}+c_{f(k,i),g(l)}-s_{f(k,i),g(l)})}{\B(a_{f,g},b_{f,g})},
\end{align*}
where $s_{f(k,i),g(l)}$ and $c_{f(k,i),g(l)}$ are respectively given in \eqref{eq:s-nobloc} and \eqref{eq:c-nobloc} below. 
These likelihoods drop into Equation~\ref{eq:Xfk_given_Rfk} and further calculations including the extension to handle labeling of unlabeled profiles are the same as for the BDP.

\subsection{NoB-LoC-CaRMa Likelihood}
\label{app:lkd-nobloc}

In NoB-LoC (transpose) the parameter $\theta^*_{f(k),g(l)}$ is no longer a scalar. There is a separate parameter for each row in a bicluster of cells $C_{f(k),g(l)}$, so we write
\[
\theta^*_{f(k),g(l)}=(\theta^*_{f(k,i),g(l)})_{i\in R_{f(k)}}.
\]
The biclustering process $R,S$ itself is just the same as BDP so we have $S_{f(k),g}=(S_{f(k),g(1)},\dots,S_{f(k),g(K_{f(k),g})})$ as the partition of $\M_g$ within the row cluster $R_{f(k)}$.

The likelihood for the mRNA profile of observations in a given fluid-type $f\in \F$, row subtype $k\in\{1,\dots, K_f\}$, marker type $g\in \F$ and marker subtype $l \in \{1, \dots, K_{f(k),g}\}$ is
\begin{align}
    p(\X_{f(k),g(l)}|\theta^*_{f(k),g(l)}, &  R_{f(k)}, S_{f(k),g(l)}) = 
    \prod_{i\in R_{f(k)}}\prod_{j\in S_{f(k),g(l)}} p(x_{i,j}|\theta_{i,j})\nonumber\\[0.1in]
     &=\prod_{i\in R_{f(k)}}
     (\theta^*_{f(k,i),g(l)})^{s_{f(k,i),g(l)}}(1-\theta^*_{f(k,i),g(l)})^{c_{f(k,i),g(l)} -s_{f(k,i),g(l)}},\label{eq:base-lkd-cs-nobloc}
\end{align}
where 
\begin{equation}\label{eq:s-nobloc}
s_{f(k,i),g(l)}=\sum_{j\in S_{f(k),g(l)}} x_{i,j}
\end{equation}
is the number of $1$'s in $\X_{f(k,i),g(l)}=(x_{i,j})_{j\in S_{f(k),g(l)}}$, and 
\begin{align}\label{eq:c-nobloc}
c_{f(k,i),g(l)} = |S_{f(k),g(l)}| 
\end{align} 
is the number of columns $j \in S_{f(k), g(l)}$. 

\subsection{NoB-LoC-CaRMa with a fixed assignment of unlabeled profiles to fluid-types}
 
  \subsubsection{NoB-LoC-CaRMa biclustering a single fluid-type with a fixed assignment partition}

Suppose the assignment partition $\V$ of unlabeled profiles $i\in \U$ to fluid-types is fixed, so for $f\in\F$ the rows in $\N_f$, and hence the set of elements partitioned by $R_f$, is fixed.
The posterior distribution of the parameters $\theta^*_f, R_f$ and $S_f$ associated with fluid-type $f$, given all data for fluid-type $f$, is
\begin{align}
    \pi(\theta^*_f, R_f, S_f  | \X_f) &\propto  \pi_R(R_f)\prod_{k=1}^{K_f}\prod_{g\in \F}\pi_S(S_{f(k),g}) \nonumber\\
    &\times \qquad \prod_{l=1}^{K_{f(k),g}} \prod_{i\in R_{f(k)}} p(\theta^*_{f(k,i),g(l)},\X_{f(k,i),g(l)}|R_{f(k)}, S_{f(k),g(l)})\nonumber
\end{align}
where for $i\in R_{f(k)}$,
\begin{align}
    p(\theta^*_{f(k,i),g(l)},\X_{f(k,i),g(l)}| & R_{f(k)}, S_{f(k),g(l)})=
     h_{f,g}(\theta^*_{f(k,i),g(l)})p(\X_{f(k,i),g(l)}|\theta^*_{f(k,i),g(l)}, R_{f(k)}, S_{f(k),g(l)})\nonumber\\
     &= \frac{(\theta^*_{f(k,i),g(l)})^{s_{f(k,i),g(l)}+a_{f,g}-1}(1-\theta^*_{f(k,i),g(l)})^{c_{f(k,i),g(l)} -s_{f(k,i),g(l)}+b_{f,g}-1}}{\B(a_{f,g},b_{f,g})}
\end{align}
The amplification probabilities $\theta^*_{f(k,i),g(l)}$ can be integrated out, which gives us
\begin{align*}
    \pi(R_f, S_{f}| X_f) 
    &\propto \pi_R(R_f)\prod_{k=1}^{K_f}\prod_{g\in \F}\pi_S(S_{f(k),g}) \prod_{l=1}^{K_{f(k),g}} p(\X_{f(k),g(l)}|R_{f(k)},S_{f(k),g(l)}),
\end{align*}
where 
\[
p(\X_{f(k),g(l)}|R_{f(k)},S_{f(k),g(l)})= \prod_{i\in R_{f(k)}} p(\X_{f(k,i),g(l)}|R_{f(k)},S_{f(k),g(l)})
\]
where for $i\in R_{f(k)}$
   \begin{align*}
   p(\X_{f(k,i),g(l)}|R_{f(k)},S_{f(k),g(l)})&=  \frac{\B(a_{f,g}+s_{f(k,i),g(l)},b_{f,g}+c_{f(k,i),g(l)}-s_{f(k,i),g(l)})}{\B(a_{f,g},b_{f,g})}.
\end{align*}
From this point on everything is the same as for the BDP. 

\section{Bayes factors}\label{app:bayes-factors}
\subsection{A variant of The Candidate's estimator}
This kind of approach is often referred to as a ``Candidate's Estimator." 
Actually, Besag's Candidate's Estimator \citep{besag89} refers to a formula for the posterior predictive distribution, but the name is still natural.  

When we just have training data
\[
\pi(R|\X,y_\T)=\prod_{f\in \F}\pi(R_f|\X_f)
\]
with 
\[
\pi(R_f|\X_f)=\frac{\pi(R_f)p(\X_f|R_f)}{p(\X_f)},
\]
the marginal likelihood for each fluid type is
\[
p(\X_f)=\frac{\pi(R_f)p(\X_f|R_f)}{\pi(R_f|\X_f)},
\]
and the overall marginal likelihood is
\[
p(\X)=\prod_{f\in\F} p(\X_f).
\]
If we can estimate $\pi(R_f|\X_f)$ reliably for any $R_f\in \Xi_{\N_f}^{J_f}$, then we obtain
\[
\widehat{p(\X_f)}=\frac{\pi(R_f)p(\X_f|R_f)}{\widehat{\pi(R_f|\X_f)}}
\]
and 
\[
\widehat{p(\X)}=\prod_{f\in\F} \widehat{p(\X_f)}.
\]


The accuracy of this method depends on estimating $\pi(R_f|\X_f)$ accurately. 
This quantity is small, even for the mode. We can do better by estimating the probability for some larger set. 
We use the HPD set for partitions. 
At level $\alpha$ this set $ \mbox{HPD}(\alpha)\subseteq \Xi^{\! *}_\N$ is defined by the conditions  
$\pi_(R|\X,y_\T)\ge \pi_(R'|\X,y_\T)$ for all $R \in \mbox{HPD}(\alpha)$ and $R' \not\in \mbox{HPD}(\alpha)$ and $\Pr(R\in \mbox{HPD}(\alpha)|\X,y_\T)=\alpha$.
We take $\alpha$ as close as possible to 0.5, for accurate estimation.

The marginal likelihood estimator based on a set,
\begin{align}
   \widehat{p(\X_f)}=\frac{\sum_{R\in \mbox{HPD}(\alpha)}\pi(R_f)p(\X_f|R_f)}{\sum_{R\in \mbox{HPD}(\alpha)} \widehat{\pi(R_f|\X_f)}} 
\end{align}
is derived in a similar way to that from a single state $R$.

The Bayes factor for two models with marginal likelihoods $p_1(\X)$ and $p_2(\X)$, with $\X$ the training data and no unlabeled profiles is estimated as
\[
\hat B_{1,2}=\prod_{f\in\F}\frac{\widehat p_1(\X_f)}{\widehat p_2(\X_f)}.
\]
This is a simple product over fluid types as $R_f$ is independent of $R_{f'}$ in the posterior $\pi_R(R|\X,y_\T)=\prod_{f\in \F}\pi_R(R_f|\X_f)$ if we analyze the training data alone (without unlabeled profiles, the set of profiles clustered by $R_f$ is $\T_f$, which is fixed), the marginal likelihood $p(\X)=\prod_{f\in \F}p(X_f)$ is a product.
\subsection{Bridge estimator}

Bridge estimation \citep{meng96} uses the identity
\[
\frac{p_{1}(\X_f)}{p_{2}(\X_f)}=\frac{E_{R_f\sim \pi_{2}}\left(p_{1}(\X_f|R_f)h(R_f)\right)}{E_{R_f\sim \pi_{1}}\left(p_{2}(\X_f|R_f)h(R_f)\right)}
\]
where $h(R_f)$ is a free function. 
The terms $\pi_1$ and $p_{1}(X_f|R_f)$ are respectively the posterior and the likelihood for NDP-CaRMa, while $\pi_2$ and $p_2$ are those for NoB-LoC-CaRMa. 
Following the recommendation in \cite{meng96}, we set
\[
h(R_f)=(\,{p_{1}(\X_f|R_f)\,p_{2}(\X_f|R_f)}\,)^{-1/2}.
\]
Let $r(X_f)={p_{1}(\X_f)}/{p_{2}(\X_f)}$. 
The estimate is
\[
\widehat{r(\X_f)}=\frac{\sum_{t=1}^T \frac{p_1(\X_f|R^{(t,2)}_f)^{1/2}}{p_2(\X_f|R^{(t,2)}_f)^{1/2}}}{\sum_{t=1}^T \frac{p_2(\X_f|R^{(t,1)}_f)^{1/2}}{p_1(\X_f|R^{(t,1)}_f)^{1/2}}}
\]
where $R_f^{(t,1)}$ are MCMC samples from $\pi_1(R_f|X_f)$ (BDP-CaRMa posterior), whereas $R_f^{(t,2)}$ are MCMC samples from $\pi_2(R_f|X_f)$ (NoB-LoC-CaRMa posterior).
Again, the Bayes factor for two models with marginal likelihoods $p_1(\X)$ and $p_2(\X)$, with $\X$ the training data and no unlabeled profiles is
\[
\hat B_{1,2}=\prod_{f\in\F} \widehat{r(\X_f)}.
\]

\section{Results}\label{sec:resultsAppdx}

\begin{figure}[htp]
        \includegraphics[width=14cm]{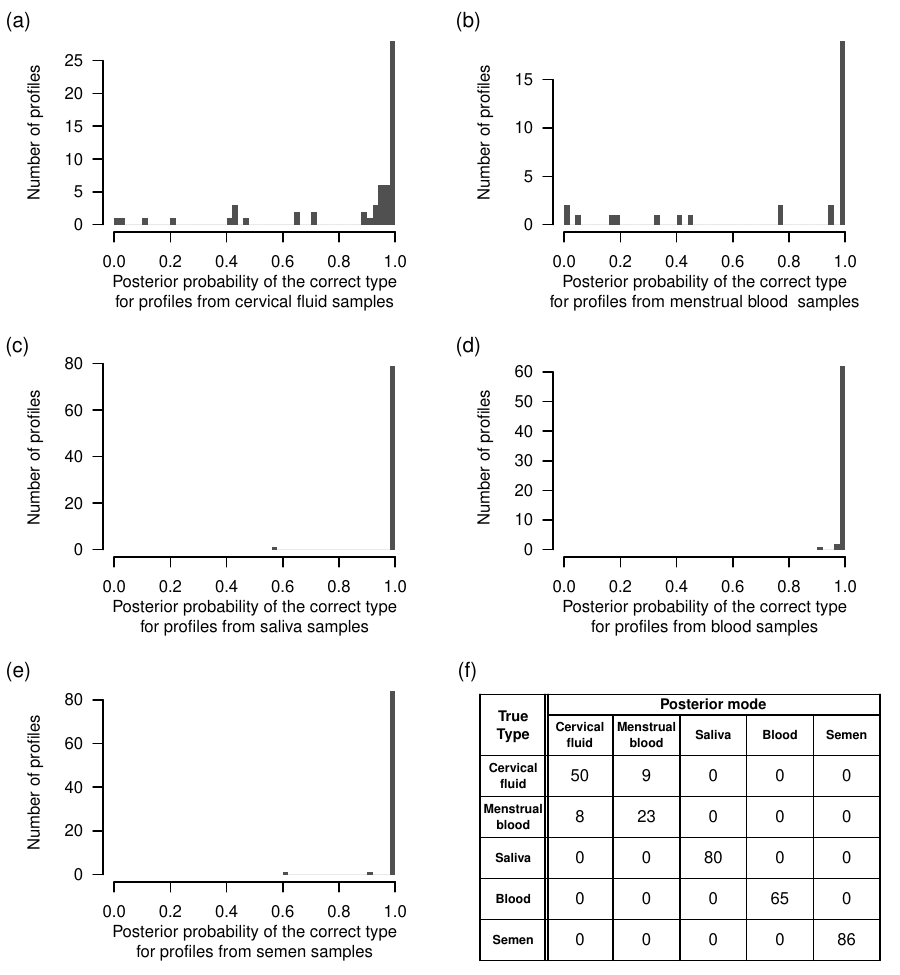}
        \centering
        \caption{(a-e) Posterior probability for the correct type estimated from the LOOCV Bayesian analysis. (f) Confusion matrix classifying using the posterior mode. This is discussed in Section~\ref{sec:loocv}.}
        \label{fig:BayesLOOCV}
\end{figure}

In section \ref{sec:loocv}, Figure \ref{fig:cutLOOCV} summarises the results of the LOOCV of the Cut-Model analysis on the training set.
For LOOCV of the Bayesian analysis on the training dataset, the summary in the same style is presented in Figure \ref{fig:BayesLOOCV}, which conveys a very similar message as Figure \ref{fig:cutLOOCV}, in particular, the confusion tables (bottom right panel) are identical in both figures.
The choice of the inference framework, Bayes or Cut-Model, does not influence the results of interest or our conclusion on the classification of fluid-type.

\section{Sensitivity analysis for bounds on partition count.}\label{app:J-sensitivity-analysis}

\subsection{Sensitivity analysis: from no biclusters to the DP}
\begin{figure}
  \caption{Comparing the posterior probabilities of the correct types estimated by Bayesian inference for the 46 mRNA profiles (a) between the default model $(J_f = 5, L_g = M_g )$ and $(J_f = 1, L_g = 1)$, and (b) between the default model and $(J_f = 15, L_g = M_g) \,\, \forall f,g \in \F$. For legibility, the positions on axes are in logit-space, while the labels represent the corresponding probability values. }
  \centering
    \includegraphics[width=\textwidth]{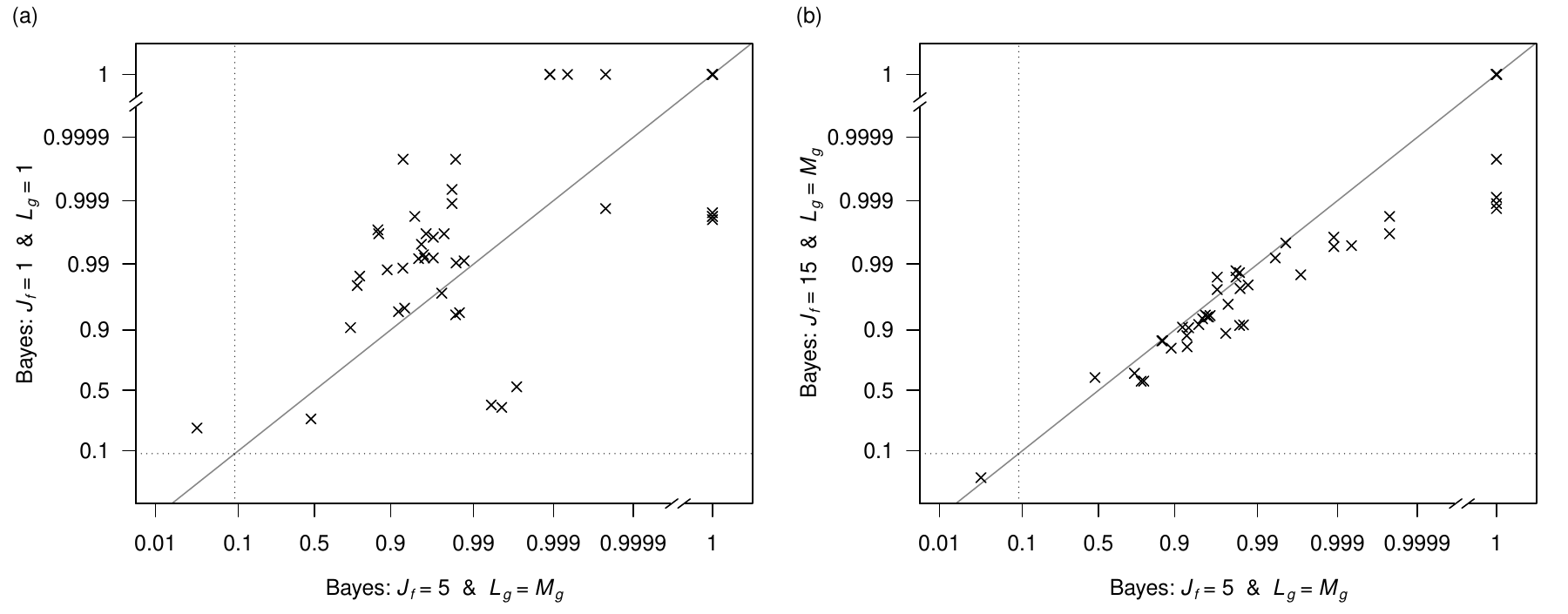}
    \label{fig:sensJCorrectTypePostDistrMain}
\end{figure}

Following model selection in Section~\ref{sec:model-selection}, we set $J_f = 5$ and $L_g = M_g$ for all $f, g \in \F$ in all following analyses.
In order to investigate whether the choice of $J_f$ has an impact on our results, we performed a sensitivity analysis with JPC on the test data with the following $(J_f, L_g)$ combinations: $(1, 1)$, $(1, M_g)$, $(10, M_g)$ and $(15, M_g)$ for all $f, g \in \F$. Results for $(1,1)$ and $(15,M_g)$ are given in Figure~\ref{fig:sensJCorrectTypePostDistrMain}. 
Results for other settings are presented in the next section. 
It is clear from the training data (Figure~\ref{fig:dataGrid}) that subtypes are a feature of the sample population for mRNA profiles. 
Figure~\ref{fig:sensJCorrectTypePostDistrMain} suggests that, when there is no biclustering, i.e., $(J_f = 1, L_g = 1) \,\, \forall f, g \in \F$, the posterior probabilities of the correct type are more dispersed than the default model.
Specifically, when the posterior probability of the correct type is approximately $0.99$, the corresponding probability estimated without biclustering can range between 0.5 and 0.9999.
Removing all biclustering tends to give posterior probabilities for correct types that are slightly higher than what we get with biclustering, as we might expect in a model with fewer parameters.
In contrast, when $(J_f, L_g) = (15, M_g) \,\, \forall f,g \in \F$, the posterior probabilities of the correct types are in fair agreement with the default model. 
There is no evidence for interaction between the choice of Bayesian or Cut-Model inference and the $(J_f, L_g)$-settings.

\begin{figure}
  \caption{Posterior distribution of the number of subtypes within each fluid-type. Each panel in the left columns is a comparison between two analyses in the Bayesian framework, while each in the right is in the Cut-Model framework.}
  \centering
    \includegraphics[width=\textwidth]{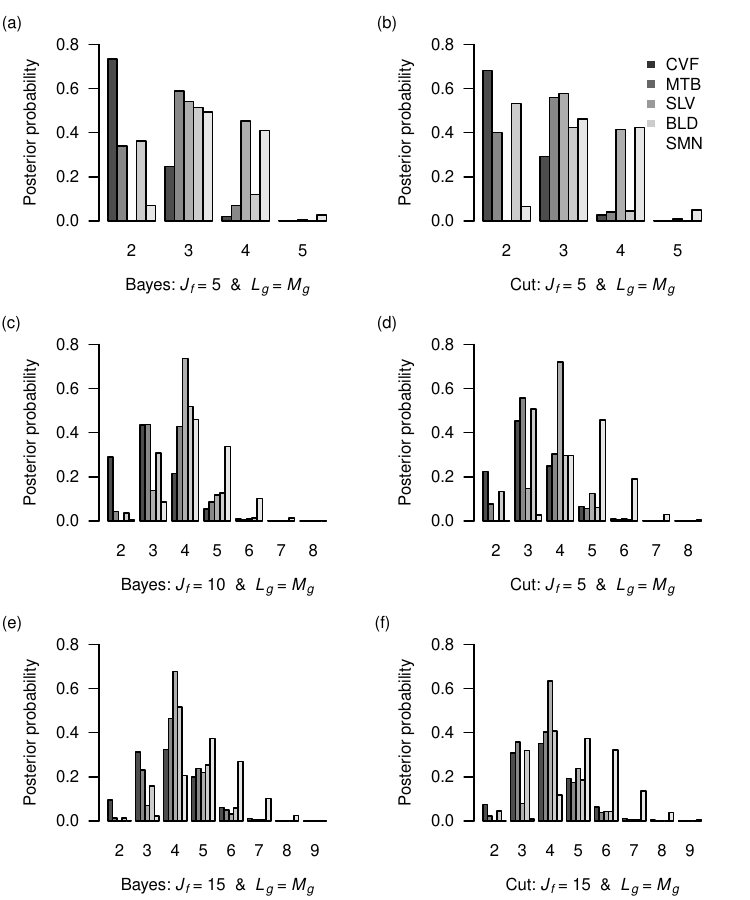}
    \label{fig:sensJSubtypeCountPostDistr}
\end{figure}

The posterior distribution of the number of subtypes within each fluid-type shifts to slightly larger numbers of subtypes, when $J_f$ is increased from 5 to 10 for all $f \in \F$ (Figure ~\ref{fig:sensJSubtypeCountPostDistr}).
Similarly, there is little change when increasing $J_f$ from 10 to 15 for all $f \in \F$. 
Considering the distributions of the numbers of subtypes when $J_f=15$, the posterior probability for six or more subtypes is small (less than 0.1) except for SMN, where it is around 0.4.
The SMN samples may require more than five subtypes to characterize the within-type heterogeneity in the marker profiles. 
This suggests there may be some value in taking a Poisson-Dirichlet-Process and estimating upper bounds separately for each fluid-type. 
However, SMN profiles have been classified with high accuracy using the default model, so this is unlikely to make much difference to the assigned class labels themselves. 

\subsection{Further Sensitivity analysis}
\begin{figure}
  \caption{Comparing the estimated posterior probabilities of the correct types of the mRNA profiles in the test set between the default model ($J_f = 5, L_g = M_g \,\, \forall f,g \in \F$) and other choices of $J_f$ and $L_g$. Each panel in the left columns is a comparison between two analyses in the Bayesian framework, while each in the right is in the Cut-Model framework. For legibility, the positions on axes are in logit-space, while the labels represent the corresponding probability values. }
  \centering
    \includegraphics[width=0.9\textwidth]{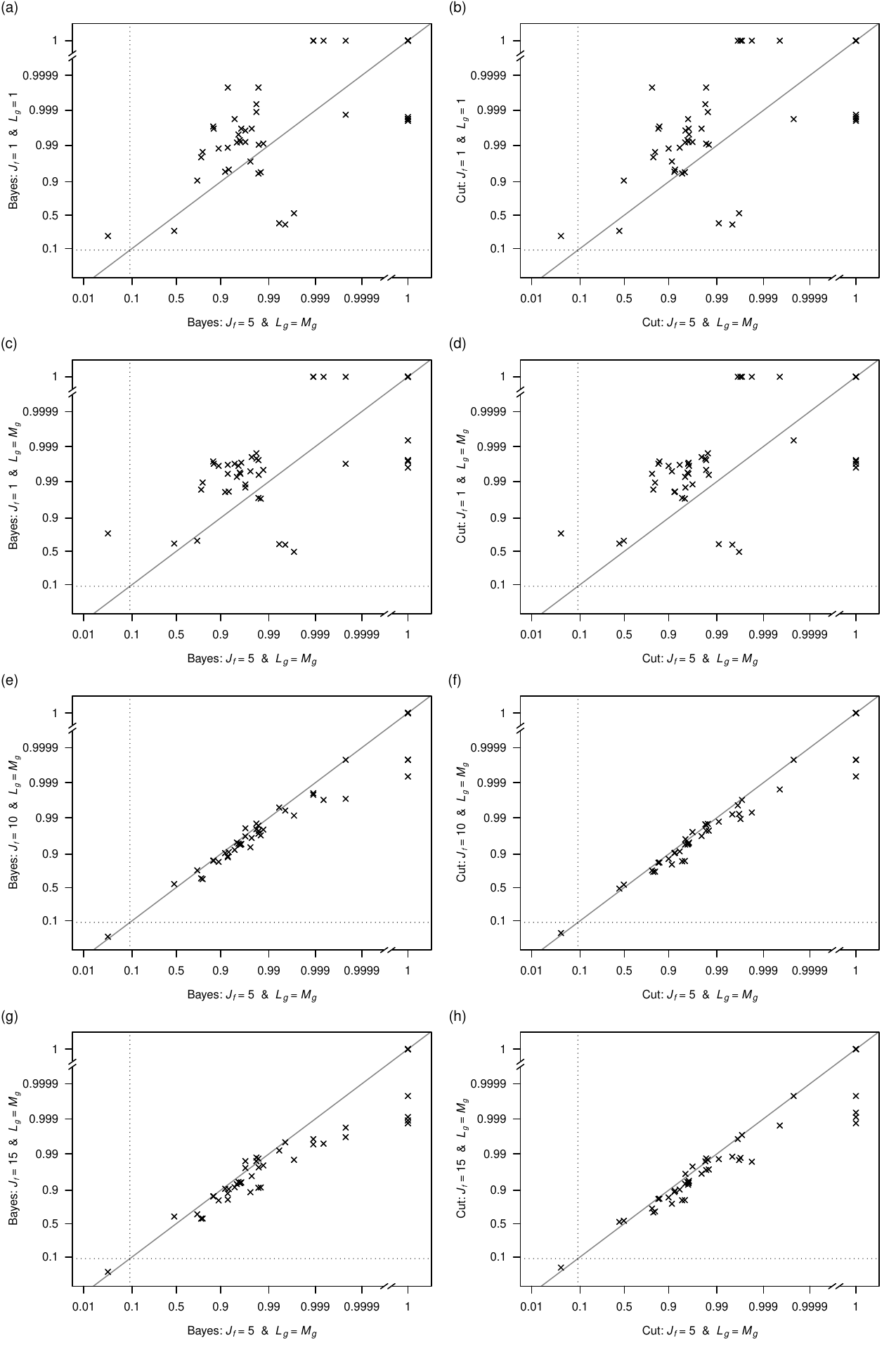}
    \label{fig:sensJCorrectTypePostDistrAll}
\end{figure}

In the previous section, we looked at sensitivity to ``extreme'' choices of $J_f$ and $L_g$. We now consider intermediate values.
Figure ~\ref{fig:sensJCorrectTypePostDistrAll} presents the comparisons in the estimated posterior probabilities of the correct types for the mRNA profiles in the test set between the default model ($J_f = 5, L_g = M_g$) and  $(J_f, L_g)$ combinations $(1, 1)$, $(1, M_g)$, $(10, M_g)$ and $(15, M_g)$ for all $f, g \in \F$.
Given a $(J_f, L_g)$ combination, the results are very similar between Bayesian and Cut-Model inferences.

From panels (a)--(d) of Figure~\ref{fig:sensJCorrectTypePostDistrAll}, there appears to be greater dispersion in the posterior probabilities of the true types for $(J_f = 1, L_g = 1)$ and $(J_f = 1, L_g = M_g)$ than the default model.
Panels (e)--(h) of Figure~\ref{fig:sensJCorrectTypePostDistrAll} show that there is generally close agreement in the posterior probabilities of the true types between the default model and $(J_f = 10, L_g = M_g)$, and between the default model and $(J_f = 15, L_g = M_g)$.
This suggests that having $J_f = 10 \text{ or } 15$ may lead to overfitting as the extra degrees of freedom add nothing of value.
These results do not depend on the choice of the inference framework.

\end{document}